\documentclass[]{JFM-FLM_Au}

\def\d{{\rm d}}
\def\i{{\rm i}}
\def\cm{{\rm cm}}
\def\Pec{{Pe}}

\def\bzero{{\boldsymbol{0}}}

\def\be{{\boldsymbol{e}}}
\def\bfv{{\boldsymbol{f}}}
\def\bh{{\boldsymbol{h}}}
\def\bk{{\boldsymbol{k}}}
\def\bmv{{\boldsymbol{m}}}
\def\bn{{\boldsymbol{n}}}
\def\bp{{\boldsymbol{p}}}
\def\br{{\boldsymbol{r}}}
\def\bu{{\boldsymbol{u}}}
\def\bv{{\boldsymbol{v}}}
\def\bx{{\boldsymbol{x}}}
\def\by{{\boldsymbol{y}}}

\def\bF{{\boldsymbol{F}}}
\def\bS{{\boldsymbol{S}}}
\def\bU{{\boldsymbol{U}}}
\def\bX{{\boldsymbol{X}}}
\def\bY{{\boldsymbol{Y}}}

\def\bA{{\mathsfbi{A}}}
\def\bG{{\mathsfbi{G}}}
\def\bI{{\mathsfbi{I}}}
\def\bQ{{\mathsfbi{Q}}}
\def\bR{{\mathsfbi{R}}}
\def\bT{{\mathsfbi{T}}}

\def\bgamma{{\boldsymbol{\gamma}}}
\def\betav{{\boldsymbol{\eta}}}
\def\bmu{{\boldsymbol{\mu}}}
\def\bnu{{\boldsymbol{\nu}}}
\def\bxi{{\boldsymbol{\xi}}}

\def\cF{{\mathcal{F}}}
\def\cK{{\mathcal{K}}}
\def\cV{{\mathcal{V}}}

\def\tc{{\tilde c}}
\def\tk{{\tilde k}}
\def\tsigma{{\tilde\sigma}}
\def\tbfv{{\tilde\bfv}}
\def\tbgamma{{\tilde\bgamma}}

\def\bbR{{\mathbb{R}}}

\def\grad{\nabla}
\def\grads{\nabla_{\rm 2D}}
\def\gradv{\nabla_{\rm 3D}}
\def\laps{\Delta_{\rm 2D}}
\def\lapv{\Delta_{\rm 3D}}

\def\ol{\overline}

\lefttitle{S. Weady}
\righttitle{Journal of Fluid Mechanics}

\title{A boundary integral formulation for microscale Marangoni surfing}

\author{Scott Weady\aff{1}}

\affiliation{
\aff{1}Center for Computational Biology, Flatiron Institute, New York, NY 10010, USA
}

\corresau{Scott Weady, \email{sweady@flatironinstitute.org}}

\begin{document}
\maketitle

\begin{abstract}
We consider a model for collections of surfactant-producing particles moving along the surface of a viscous fluid under their self-generated Marangoni flows. Such systems are challenging due to their nonlocal nature: particle motion is coupled to Marangoni stresses distributed across the entire fluid surface and to the resulting flows in the bulk. In the low Reynolds and P\'eclet number limit, we reduce these nonlocal effects to a system of integral equations supported solely on the particles and their boundaries. This formulation is not only computationally tractable, but also provides a framework for systematically coarse-graining many-particle systems. Numerical simulations, performed with a custom spectral method, show collective dynamics are highly sensitive to shape. In particular, among elliptical particles, we find prolate particles tend to attract while oblate particles tend to repel, and these geometric details control the stability of polarized populations. We rationalize these behaviors through a far-field expansion of the boundary integral equations and its corresponding reduced-order model.
\end{abstract}

\section{Introduction}

The Marangoni effect describes the flow that develops when a fluid interface possesses gradients in surface tension. Such gradients may arise, for example, from a non-uniform distribution of surfactant molecules, temperature, or electric charge \citep{scriven1960marangoni,manikantan2020surfactant}. Various organisms from micron to millimetric scales exploit the Marangoni effect for locomotion \citep{bush2006walking}. For example, the aquatic insect {\em Microvelia} can secrete surfactant to propel itself towards land, reaching speeds of tens of body lengths per second \citep{andersen1982semiaquatic}. At the microscale, surfactant-producing microorganisms such as {\em P. aeruginosa} and {\em B. subtilis} exhibit collective motion as a consequence of Marangoni flows, and these collective motions can enhance the rate of colonization \citep{angelini2009bacillus,fauvart2012surface}. Such organisms are examples of ``Marangoni surfers,'' which refer to particles that self-propel along a fluid interface due to self-generated surface tension gradients \citep{fei2017active}.

The potential for surface tension-based propulsion has been known for centuries \citep{tomlinson1860motions,rayleigh1889measurements}, and has become of recent interest not only for its relevance to biological locomotion, but also for its practical potential in directed transport and self-assembly \citep{dietrich2020microscale,kwak2021marangoni,timm2021remotely,zhang2025edible}. At the single particle level, exact solutions for ellipsoidal surfers in the low Reynolds and P\'eclet number limit show that accounting for viscous stresses induced by bulk Marangoni flows is necessary for predicting the speed of propulsion \citep{lauga2012viscous,masoud2014reciprocal,wurger2014thermally}. Analytical and numerical solutions that incorporate confinement, geometry, surfactant degradation, and/or advective transport identify increasingly rich phenomenology, with possible flow reversal, arrest, and translational-rotational coupling \citep{sur2019translational, jafari2020forward,crowdy2021viscous,zhu2023self,kang2026convective}.

While single Marangoni surfers have been thoroughly studied, interactions among multiple surfers are less understood. Experimental work in this direction has primarily focused on camphor boats in strong confinement, which show surface tension and hydrodynamic interactions can lead to highly ordered states and swarms \citep{kohira2001synchronized,soh2008dynamic,heisler2012swarming}. From a modeling perspective, multi-particle systems are challenging not only due to their multi-component geometries, but also due to the nonlocal coupling induced by the fluid substrate, an important effect which is often neglected in reduced models \citep{boniface2021role}. Analytical solutions have been found in simplified two-dimensional settings, which show collectively generated Marangoni stresses can enhance group propulsion \citep{crowdy2020collective}. In three dimensions, numerical simulations of point-like pusher particles show that Marangoni stresses can stabilize periodic trajectories \citep{mottammal2021collective}. Continuum models, derived under well-separated assumptions, also demonstrate collective behavior, including apparent finite time singularities in the particle density \citep{masoud2014collective}. These works provide insight into collective behavior, but do not resolve the coupling between geometry, Marangoni stresses, and fluid flow. How this coupling governs collective dynamics remains an open question.

In this work, we develop a boundary integral formulation for analyzing the collective dynamics of Marangoni surfers. We consider the low Reynolds and P\'eclet number limit, which is a plausible approximation for surfactant-producing microorganisms such as {\it P. aeruginosa} or {\it B. subtilis}. A key insight of this formulation is that, in this limit, the net effect of the Marangoni stress, which is distributed across the entire fluid interface, can be evaluated in terms of a density supported on the particle boundaries. This formulation not only simplifies discretization of the system geometry, but also admits asymptotic expansions that can be used to systematically coarse-grain the dynamics of many well-separated surfers. Numerical solution of the boundary integral equations, enabled by an efficient spectral method, and analysis of the coarse-grained theory identifies a range of collective behaviors, including aggregation, repulsion, and enhanced spreading.

The paper is organized as follows. In sections \ref{sec:model}--\ref{sec:numerical-method} we formulate the model, derive the boundary integral equations, and describe a numerical method for elliptical particles. Sections \ref{sec:single} and \ref{sec:far-field} involve analysis of the boundary integral equations, including the calculation of the propulsion speed of an elliptical particle and the far-field approximation to the flow induced by a collection of well-separated particles. Finally, we simulate the model in sections \ref{sec:pair} and \ref{sec:collective} and rationalize our observations using the far-field approximation.

\section{Mathematical model}\label{sec:model}

In this section we describe a mathematical model for a collection of Marangoni surfers. The model follows that of \cite{lauga2012viscous} in which a surfactant, insoluble with the bulk fluid, diffuses along the surface of a semi-infinite fluid volume. For simplicity we neglect capillary interactions and assume the surface remains planar. While we focus on this particular model for consistency with prior work, the approach taken here applies to other models that may include sublimation, bulk diffusion, or flux boundary conditions; further discussion can be found in appendices \ref{app:bulk-diffusion} and \ref{app:sublimation}. 

\begin{figure}
    \includegraphics[width=\linewidth]{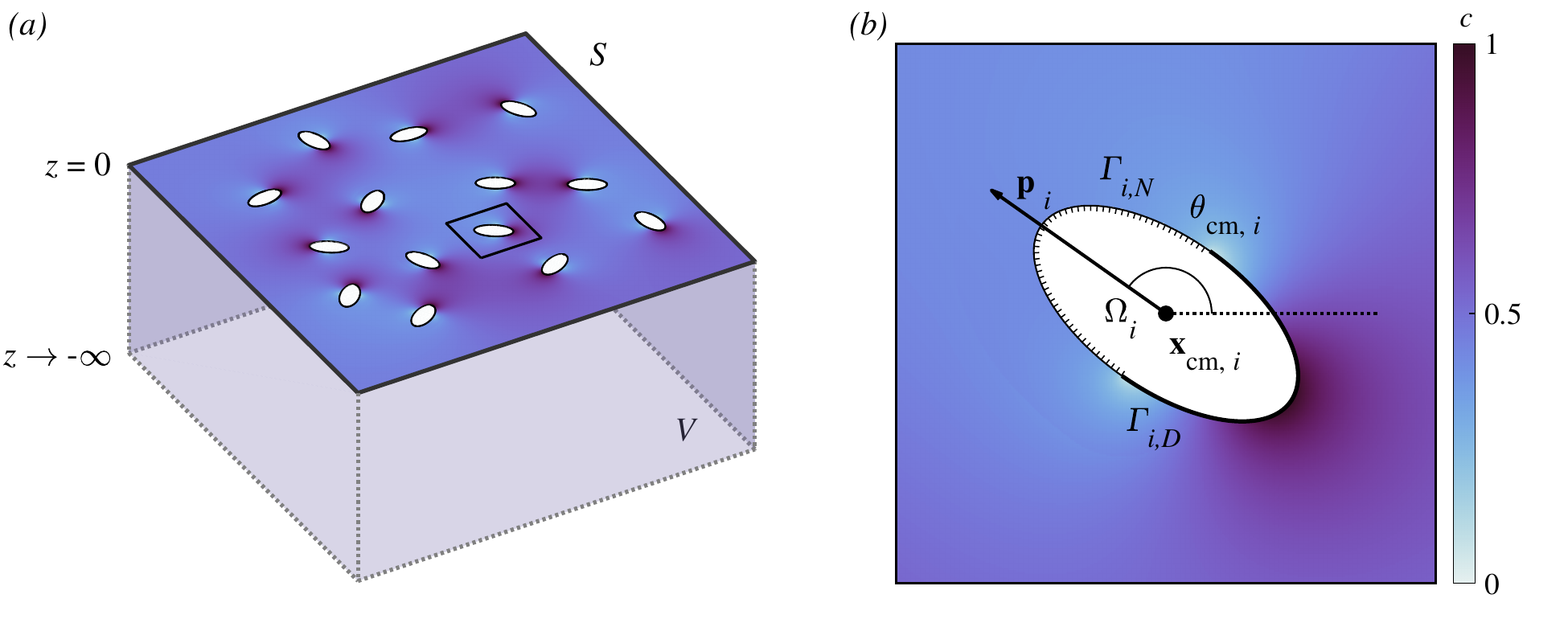}
    \caption{Schematic of the mathematical model. {\em(a)} A collection of rigid surfactant-producing particles (white) reside on the surface $S$ of a semi-infinite fluid volume $V$. Surfactant (color) generated on the particle boundaries diffuses along the surface, generating Marangoni stresses. {\it(b)} Each particle produces surfactant along a portion of its boundary, $\Gamma_{i,D}$, and is impermeable along the remaining portion, $\Gamma_{i,N}$.}\label{fig:schematic}
\end{figure}

We consider $N$ particles residing on the surface $S = \{\bX = (x,y,0) \in \bbR^3\}$ of a semi-infinite fluid volume $V = \{ \bX = (x,y,z) \in \bbR^3 : z < 0 \}$. The configuration is described by the particle center of masses $\bx_{\cm,i} \in \bbR^2$ and orientation angles $\theta_{\cm,i} \in [0,2\pi)$ for $i = 1,\ldots,N$. The solid domain of particle $i$ is given by the set $\Omega_i\subset S$ with boundary $\Gamma_i = \partial\Omega_i$. The collection of particle domains is denoted by $\Omega = \bigcup_{i=1}^N \Omega_i$ and the collection of particle boundaries by $\Gamma = \bigcup_{i=1}^N \Gamma_i$. The portion of the surface exterior to the particles is therefore $S\setminus\ol{\Omega}$. A schematic of the system is shown in figure \ref{fig:schematic}.

\subsection{Velocity}

The bulk velocity $\bU : V \to \bbR^3$ and pressure $P : V \to \bbR$ satisfy the incompressible Navier-Stokes equations,
\begin{align}
    \rho\left(\frac{\partial\bU}{\partial t} + \bU\cdot\gradv\bU\right) = -\gradv P + \mu\lapv \bU &\quad {\rm in} ~ V,\\
    \gradv \cdot\bU = 0 &\quad {\rm in} ~ V,
\end{align}
where $\rho$ is the fluid density and $\mu$ is the dynamic viscosity, both assumed constant, $\gradv = (\partial_x,\partial_y,\partial_z)$ is the three-dimensional gradient, and $\lapv = \partial_x^2 + \partial_y^2 + \partial_z^2$ is the three-dimensional Laplacian.

\subsection{Surfactant}

On the surface exterior to the particles, the surfactant concentration $c$ satisfies the advection-diffusion equation
\begin{equation}
    \frac{\partial c}{\partial t} + \grads\cdot(\bu c) = D\laps c \quad {\rm on} ~ S\setminus\ol{\Omega},
\end{equation}
where $D$ is a diffusion coefficient, $\grads = (\partial_x,\partial_y)$ is the two-dimensional gradient, and $\laps = \partial_x^2 + \partial_y^2$ is the two-dimensional Laplacian. Variations in surfactant concentration modify the surface tension $\gamma(c)$, which in general may be a nonlinear function of $c$. Here we take the simple linear equation of state $\gamma(c) = \gamma_0 - \gamma_1c$, where $\gamma_1$ is a positive constant \citep{manikantan2020surfactant}.

\subsection{Boundary conditions}

On the fluid surface exterior to the particles, we have the Marangoni boundary condition
\begin{equation}
    \mu\frac{\partial(U_x,U_y)}{\partial z} = \grads\gamma(c) \quad {\rm on} ~ S\setminus\ol{\Omega}.
\end{equation}
We assume the particles move tangentially along the surface, which gives the no-slip condition,
\begin{equation}
    (U_x, U_y) = \bu_{\cm,i} + \omega_{\cm,i} (\bx-\bx_{\cm,i})^\perp \quad {\rm on} ~ \Omega_i,
\end{equation}
where $\bu_{\cm,i}$ is the translational velocity of particle $i$, $\omega_{\cm,i}$ is the angular velocity, and $\bx^\perp = (-y,x)$ denotes rotation by $\pi/2$. The surface is also assumed to remain flat everywhere,
\begin{equation}
    \bU\cdot\be_z = 0 \quad {\rm on} ~ S,
\end{equation}
and in the far-field we assume the velocity decays to zero,
\begin{equation}
    \bU\to\bzero \quad {\rm as} ~ |\bX|\to\infty.
\end{equation}

We consider particles that have a prescribed surfactant concentration $g_i$ on a subset of the boundary $\Gamma_{i,D}$ and are impermeable on another portion $\Gamma_{i,N} = \Gamma_i \setminus \ol{\Gamma_{i,D}}$. We denote $\Gamma_D = \bigcup_{i=1}^N \Gamma_{i,D}$ and $\Gamma_N = \bigcup_{i=1}^N \Gamma_{i,N}$ as the collection of all such boundaries. Defining $g : \Gamma_D \to \bbR$ such that $g\rvert_{\Gamma_{i,D}} = g_i$, the mixed boundary condition is
\begin{align}
    c = g &\quad {\rm on} ~ \Gamma_D,\\
    \frac{\partial c}{\partial n} = 0 &\quad{\rm on} ~ \Gamma_N.
\end{align}
Finally, in the far-field we require the concentration to be bounded,
\begin{equation}
    |c| < \infty \quad {\rm as} ~ |\bx|\to\infty.
\end{equation}
At finite P\'eclet numbers, the far-field concentration can be prescribed, however in the low P\'eclet number limit, which we will consider here, the concentration is determined by the requirement of zero net flux on the collection of particle boundaries. 

\subsection{Rigid body dynamics}

The no-slip condition induces a force density $\bfv : \Omega \to \bbR^2$, corresponding to the tangential force the particle exerts on the fluid. For particles under no external forces and torques, in terms of the restriction $\bfv_i = \bfv\rvert_{\Omega_i}$ to each particle $i = 1,\ldots, N$, the center of mass dynamics are given by Newton's laws,
\begin{align}
    M_i\frac{\d^2 \bx_{\cm,i}}{\d t^2} = -\left[\int_{\Omega_i} \bfv_i ~ \d S + \gamma_1\int_{\Gamma_i} c \hat\bn_i ~ \d \Gamma\right],\\    
    I_i\ddot\theta_{\cm,i} = -\left[\int_{\Omega_i} (\bx-\bx_{\cm,i})^\perp\cdot\bfv_i ~ \d S + \gamma_1\int_{\Gamma_i} c [(\bx-\bx_{\cm,i})^\perp\cdot\hat\bn_i] ~ \d \Gamma\right],
\end{align}
where $M_i$ is the mass of particle $i$, $I_i$ is the moment of inertia and $\hat\bn_i$ is the outward normal to the particle. The first terms on the right hand side describe fluid forces and torques while the second terms describe forces and torques arising from surface tension.

\subsection{Non-dimensionalization}\label{sec:nondimensionalization}

We choose a characteristic concentration scale $c_* = \max_{\Gamma_D} |g|$, length scale $\ell_* = (|\Omega_i|/\pi)^{1/2}$, time scale $t_* = (\mu\ell_*)/(\gamma_1 c_*)$, velocity scale $u_* = \ell_*/t_*$, and pressure scale $P_* = (\gamma_1 c_*) / \ell_*$. (Note that these scales provide a rough order of magnitude estimate; the actual particle velocity and corresponding time scale are determined {\em a posteriori}.) Assuming the Reynolds number $\Rey = (\rho \ell_*u_*)/\mu$ is small, which holds for microorganisms of typical size $\ell_* \sim 1-10~\mu{\rm m}$ moving at speeds of order $u_* \sim 1-10~\mu{\rm m/s}$ in water, the bulk velocity satisfies the Stokes equations
\begin{align}
    -\gradv' P' + \lapv' \bU' = \bzero &\quad {\rm in} ~ V',\\
    \gradv'\cdot\bU' = 0 &\quad {\rm in} ~ V',
\end{align}
with boundary conditions
\begin{align}
    \frac{\partial(U_x', U_y')}{\partial z'} + \grads' c' = \bzero &\quad {\rm on} ~ S'\setminus\ol{\Omega}',\\
    (U_x',U_y') - \bu_{\cm,i}' - \omega_{\cm,i}'(\bx'-\bx_{\cm,i}')^\perp = \bzero &\quad{\rm on} ~ \Omega_i',\\
    \bU'\cdot\be_z = 0 &\quad {\rm on} ~ S',
\end{align}
where dimensionless variables are denoted by primes. In this same limit, the force and torque balances become constraints,
\begin{align}
\int_{\Omega_i'} \bfv_i' ~ \d S' + \int_{\Gamma_i'} c' \hat\bn_i ~ \d\Gamma' &= \bzero,\\
\int_{\Omega_i'} (\bx'-\bx_{\cm,i}')^{\perp}\cdot\bfv_i' ~ \d S' + \int_{\Gamma_i'} (\bx'-\bx_{\cm,i}')^\perp\cdot c'\hat\bn_i ~ \d\Gamma' &= 0,
\end{align}
and the translational and rotational velocities are their associated unknowns. We further assume that the P\'eclet number $\Pec = (\ell_*u_*)/D$ is small. For $D \sim 10^{-10}$--$10^{-9}~{\rm m^2/s}$, which is typical of small surfactants in water, the microorganismal scales above give $\Pec \sim 10^{-3}$--$10^{0}$, so the assumption is appropriate at the smaller and slower end of this range. Under this assumption, the surfactant concentration satisfies the Laplace equation
\begin{align}
    -\laps'c' = 0 &\quad {\rm on} ~ S'\setminus\ol\Omega',\\
    c' = g' &\quad {\rm on} ~ \Gamma_D',\\
    \frac{\partial c'}{\partial n'} = 0 &\quad{\rm on} ~ \Gamma_N',\\
    |c'| < \infty &\quad {\rm as} ~ |\bx'|\to\infty,
\end{align}
where $g' = g/c_*$. The only free parameters remaining are the dimensionless boundary surfactant concentration $g'$ and the particle geometry. In the rest of the manuscript we assume all variables are dimensionless and omit primes. We will also use the shorthand $\grad := \grads$ for the surface gradient and $\Delta := \laps$ for the surface Laplacian.

\subsection{Parameterization}

While the formulation is agnostic to the specific particle shape, our computations focus on elliptical particles with principal axes of half length $a$ and $b$. (Note we do not assume the convention $a \ge b$.)  We thus parameterize each $\Gamma_i$ by $\bgamma_i(s_i) = \bx_{\cm,i} + \bQ(\theta_{\cm,i})\cdot\bgamma_0(s_i)$ for $s_i \in [-\pi/2, 3\pi/2)$, where $\bgamma_0(s) = (a\cos s, b\sin s)$ is the reference parameterization of an ellipse centered at the origin and oriented with its principal axis of half length $a$ directed along the $x$-axis, and $\bQ(\theta)$ is the rotation matrix
\begin{equation}
\bQ(\theta) = \begin{pmatrix} \cos\theta & -\sin\theta \\ \sin\theta & \cos\theta \end{pmatrix}.
\end{equation}
We define the Dirichlet boundary $\Gamma_{i,D}$ by $s_i \in (\pi/2, 3\pi/2)$ and the Neumann boundary $\Gamma_{i,N}$ by $s_i \in (-\pi/2,\pi/2)$. Along the Dirichlet boundary, we choose the surfactant profile $g_i(s_i) = -\cos s_i$, which corresponds to a peak at the tail that tends to zero at the intersection $\Gamma_{i,D}\cap\Gamma_{i,N}$. Similarly, the interior $\Omega_i$ is parameterized by $\bx_i(\bxi_i) = \bx_{\cm,i} + \bQ(\theta_{\cm,i})\cdot\bx_0(\bxi_i)$ for $\bxi_i \in D$ in the unit disk $D$, where $\bx_0(\bxi) = (a\xi_1,b\xi_2)$ is the reference parameterization.

\section{Boundary integral formulation}\label{sec:bie}

In the limit $\Rey,\Pec \ll 1$, the equations are linear, constant coefficient, and elliptic and are thus amenable to boundary integral methods. Exploiting these properties, here we reduce the full coupled surface-bulk problem to a system of integral equations supported on the particles and their boundaries. Analogous formulations for bulk diffusion and surfactant sublimation are given in appendices \ref{app:bulk-diffusion} and \ref{app:sublimation}.

\subsection{Velocity}

We represent the bulk velocity field as
\begin{equation}
    \bU(\bX) = \int_S \bG_{\rm 3D}(\bX - \bY) \cdot \bF(\bY) ~ \d S_{\bY} \quad \bX \in V,\label{eq:U=S[F]}
\end{equation}
where $\bG_{\rm 3D}(\bR) = (\bI_{\rm 3D}+\hat\bR\hat\bR)/(8\pi R)$ is the Stokeslet. This representation automatically satisfies the Stokes equations in $V$ along with the decay condition $\bU\to\bzero$ as $|\bX|\to\infty$. Using the traction boundary conditions, the Lorentz reciprocal theorem, and the no-penetration condition $\bU\cdot\be_z = 0$, the force density has the explicit form \citep{pozrikidis1992boundary},
\begin{equation}
    \bF = \begin{pmatrix} 2(\chi_\Omega \bfv - \chi_{S\setminus\ol\Omega} \grad c) \\ 0 \end{pmatrix},
\end{equation}
where $\chi_B$ is the characteristic function, which provides the natural zero extension
\begin{equation}
    (\chi_B \nu)(\bx) = \begin{cases} \nu(\bx) & \bx \in B, \\ 0 & \bx \in S \setminus \ol{B}.\end{cases}
\end{equation}
Restricting to the surface and using continuity of the representation \eqref{eq:U=S[F]} up to the boundary gives an expression for the surface velocity
\begin{equation}
    \bu = \bS_\Omega[\bfv] - \bS_{S\setminus\ol\Omega}[\grad c] \quad {\rm on} ~ S,
\end{equation}
where we defined the surface-to-surface single layer potential supported on the set $B$,
\begin{equation}
    \bS_B[\bnu](\bx) := \int_B \bG(\bx-\by) \bnu(\by) ~ \d S_{\by},
\end{equation}
with $\bG(\br) = (\bI + \hat\br\hat\br)/(4\pi r)$ twice the tangential projection of the Stokeslet. Note that $\bG$ is a $2\times 2$ matrix-valued function.

\subsection{Surfactant}

We represent the surfactant concentration field as 
\begin{equation}
    c = c_\infty + \cV_\Gamma[\sigma] \quad {\rm on} ~ S\setminus\ol{\Omega},\label{eq:c=V[sigma]}
\end{equation}
where $c_\infty$ is an {\em a priori} unknown constant and
\begin{equation}
    \cV_\Gamma[\sigma](\bx) = \int_\Gamma G(\bx - \by) \sigma(\by) ~ \d\Gamma_{\by}
\end{equation}
is the single layer potential of the two-dimensional Laplace equation with $G(\br) = -(1/2\pi)\log|\br|$ its fundamental solution, i.e. $-\Delta G = \delta(\br)$. Letting $\sigma_i = \sigma\rvert_{\Gamma_i}$, the densities $\sigma_i$ and far-field concentration $c_\infty$ can be determined by substituting the single layer potential into the surfactant boundary conditions, resulting in the mixed integral equation
\begin{align}
    c_\infty + V_{\Gamma_i}[\sigma_i] + \sum_{j\neq i}^N \cV_{\Gamma_j} [\sigma_j] = g_i &\quad {\rm on} ~ \Gamma_{i,D},\\
    -\frac{\sigma_i}{2} + K'_{\Gamma_i}[\sigma_i] + \sum_{j\neq i}^N \cK'_{\Gamma_i,\Gamma_j}[\sigma_j] = 0 &\quad {\rm on} ~ \Gamma_{i,N},\\
    \int_\Gamma \sigma ~ \d\Gamma = 0 &,
\end{align}
where
\begin{equation}
    V_{\Gamma_i}[\sigma_i](\bx) = -\frac{1}{2\pi} \int_{\Gamma_i} \log |\bx-\by| \sigma_i(\by) ~ \d\Gamma_{\by} \quad \bx \in \Gamma_i
\end{equation}
is the self-evaluated single layer potential,
\begin{equation}
    K'_{\Gamma_i}[\sigma_i](\bx) = -\frac{1}{2\pi} \int_{\Gamma_i} \frac{\hat\bn_i(\bx)\cdot(\bx-\by)}{|\bx-\by|^2} \sigma_i(\by) ~ \d\Gamma_{\by} \quad \bx \in \Gamma_{i,N}
\end{equation}
is the adjoint of the self-evaluated double layer potential, and
\begin{equation}
    \cK'_{\Gamma_i,\Gamma_j}[\sigma_j](\bx) = \hat\bn_i(\bx) \cdot \grad \cV_{\Gamma_j}[\sigma_j](\bx) \quad \bx \in \Gamma_{i,N}
\end{equation}
is the normal derivative, with respect to the outward normal $\hat\bn_i$ on the particle, of the Laplace single layer potential on particle $i$ due to the presence of particle $j$. The term $-\sigma_i/2$ appearing in the Neumann condition arises from the jump relation of the adjoint double layer potential when approaching the boundary from the exterior. The constraint $\int_\Gamma \sigma ~ \d\Gamma = 0$, which is integrated over all particles, ensures $c$ is bounded at infinity.

\subsection{Particle-supported formulation}

The system of equations for the surfactant densities $\sigma_i$ and rigid body forces and velocities $\bfv_i, \bu_{\cm,i}, \omega_{\cm,i}$ is closed and can in principle be solved numerically. However, the Marangoni stress is distributed on the surface exterior to the particle, $S\setminus\ol{\Omega}$, which is of infinite extent and is therefore computationally demanding. An analogous integral appears in the reciprocal formulation of a Marangoni surfer \citep{masoud2014reciprocal} and can be computed analytically using known rigid body solutions in structured geometries. Here we reduce this contribution to an integral supported on $\Gamma$. A key observation is that the single layer representation of the surfactant concentration \eqref{eq:c=V[sigma]} is continuous and harmonic, and can be used to extend $c$ to all of $S$. Importantly, this extension allows for natural use of the Fourier transform.

Denoting the extension by $\tc$, we may write
\begin{equation}
\begin{aligned}
    \bu &= \bS_\Omega[\bfv + \grad\tc] - \bS_S[\grad\tc]    
    \\ & =
    \bS_\Omega[\tbfv] - \bS_S[\grad\tc],
\end{aligned}
\end{equation}
where $\tbfv = \bfv + \grad \tc$. This effective force density has a concise interpretation. In particular, integrating over ${\Omega_i}$, using the definition $\tbfv_i = \bfv_i + \grad\tc$ and applying the divergence theorem, we find $\int_{\Omega_i} \tbfv_i ~ \d S = \bzero$ and $\int_{\Omega_i} (\bx-\bx_{\cm,i})^\perp\cdot\tbfv_i ~ \d S = 0$, which are effective force- and torque-free conditions.

We denote the Fourier transform by $\cF[\nu](\bk) = \int_{\bbR^2} \nu(\bx) e^{-\i \bk \cdot \bx} ~ \d S_{\bx}$. Using $\cF[G] = 1/k^2$ and the convolution theorem, we have
\begin{equation}
    \begin{aligned}
        \grad \tc &= \grad (G * (\delta_\Gamma \sigma)) \\
        &= 
        \cF^{-1}\left[\frac{\i\bk}{k^2} \widehat{(\delta_\Gamma \sigma)}\right],
    \end{aligned}
\end{equation}
where $\delta_\Gamma$ denotes the Dirac delta distribution supported on $\Gamma$. Moreover, using $\cF[\bG] = \bI/k -\bk\bk/(2k^3)$, we find
\begin{equation}
    \cF[\bG * \grad\tc] = \frac{\i\bk}{2k^3}\cF[\delta_\Gamma \sigma].
\end{equation}
The symbol $1/k^3$, which formally corresponds to an inverse fractional Laplacian $(-\Delta)^{-3/2}$, does not have a classical inverse Fourier transform and must be understood as a distribution \citep{hsiao2021boundary}. Explicitly, the inverse transform can be determined from $\cF^{-1}[1/k^3] = (-\Delta)^{-1}[\cF^{-1}[1/k]] = (-\Delta)^{-1}[1/(2\pi|\br|)]$, which gives $\cF^{-1}[1/k^3] = -|\bx|/2\pi$ up to an additive constant. Applying the convolution theorem again, we arrive at an integral representation for the surface velocity,
\begin{equation}
    \bu = \bS_\Omega[\tbfv] + \grad  B_\Gamma[\sigma],
\end{equation}
where the integral operator
\begin{equation}
    B_\Gamma[\sigma](\bx) = \frac{1}{4\pi} \int_\Gamma |\bx-\by|\sigma(\by) ~ \d \Gamma_{\by}
\end{equation}
captures nonlocal effects of the bulk Marangoni flow. (The bulk velocity can be similarly represented; see appendix \ref{app:bulk-flow}.) Note that, because we are taking a gradient, the constant does not affect the velocity and can be taken to be zero without loss of generality. It is often useful to commute the gradient with the integral so that $\grad B_\Gamma$ may also be evaluated as
\begin{equation}
    \grad B_\Gamma[\sigma] = \frac{1}{4\pi} \int_\Gamma \frac{\bx - \by}{|\bx - \by|} \sigma(\by) ~ \d\Gamma_{\by}.
\end{equation}

\subsection{The boundary integral equations}

Evaluating the surface velocity on each $\Omega_i$ and using the no-slip condition yields a closed system of equations for the rigid body velocities $(\bu_{\cm,i},\omega_{\cm,i})$, force densities $\tbfv_i$, surfactant densities $\sigma_i$, and far-field concentration $c_\infty$,
\begin{align}
    \bu_{\cm,i} + \omega_{\cm,i}(\bx-\bx_{\cm,i})^\perp - \left(\sum_{j=1}^N \bS_{\Omega_j}[\tbfv_j] + \grad B_{\Gamma_j}[\sigma_j]\right) = \bzero &\quad{\rm on} ~ \Omega_i,\label{eq:BIE_1}\\
    c_\infty + V_{\Gamma_i}[\sigma_i] + \sum_{j\neq i}^N \cV_{\Gamma_j} [\sigma_j] = g_i &\quad {\rm on} ~ \Gamma_{i,D},\\
    -\frac{\sigma_i}{2} + K'_{\Gamma_i}[\sigma_i] + \sum_{j\neq i}^N \cK'_{\Gamma_i,\Gamma_j}[\sigma_j] = 0 &\quad {\rm on} ~ \Gamma_{i,N},\\
    \int_{\Omega_i} \tbfv_i ~ \d S = \bzero,\\
    \int_{\Omega_i} (\bx-\bx_{\cm,i})^\perp\cdot\tbfv_i ~ \d S = 0,&\\
    \sum_{i=1}^N \left(\int_{\Gamma_i} \sigma_i ~ \d \Gamma\right) = 0 &.\label{eq:BIE_N}
\end{align}
This formulation is supported only on the particles and their boundaries and is thus amenable to numerical solution. This is nonetheless nontrivial, and we describe an accurate and efficient method for its discretization and solution in the following section.

\section{Numerical method}\label{sec:numerical-method}

There are several computational challenges involved in solving the boundary integral equations \eqref{eq:BIE_1}--\eqref{eq:BIE_N}. In particular, the force density $\tbfv$ generically exhibits an inverse square-root singularity in the distance to $\Gamma$, and the density $\sigma$ has an expansion in half-integer powers, possibly with an inverse square-root, in the distance to the Dirichlet-Neumann junction $\ol{\Gamma_D}\cap\ol{\Gamma_N}$ \citep{akhmetgaliyev2017regularized}. These singularities are typical of mixed boundary value problems, and their known form can be exploited to develop accurate methods of solution. Here we outline an efficient spectral method for elliptical particles that resolves these singularities. A more detailed description is provided in appendices \ref{app:stokes} and \ref{app:laplace}.

\subsection{Representation of the force density}

The Stokes problem for the force densities and rigid body velocities can be written as the block system
\begin{equation}
\begin{pmatrix}
-\bS_\Omega & \bR_\Omega \\
\bR_\Omega^* & \bzero
\end{pmatrix}
\begin{pmatrix} \tbfv \\ \begin{pmatrix} \bu_\cm \\ \omega_\cm \end{pmatrix} \end{pmatrix}
=
\begin{pmatrix} \grad B_\Gamma[\sigma] \\ \bzero \end{pmatrix},\label{eq:stokes-system}
\end{equation}
where $\bS_\Omega$ is the $N\times N$ block operator with entries $(\bS_\Omega)_{ij}[\tbfv_j] = \bS_{\Omega_j}[\tbfv_j]\rvert_{\Omega_i}$ and $\tbfv = (\tbfv_1,\ldots,\tbfv_N)$, $\bu_\cm = (\bu_{\cm,1},\ldots,\bu_{\cm,N})$, and $\omega_\cm = (\omega_{\cm,1},\ldots,\omega_{\cm,N})$. The operator $\bR_\Omega$ maps rigid body velocities to the corresponding velocity field on each particle, namely $(\bR_\Omega(\bu_\cm,\omega_\cm))_i = \bu_{\cm,i} + \omega_{\cm,i}(\bx-\bx_{\cm,i})^\perp$ on $\Omega_i$, and its adjoint $\bR_\Omega^*$ gives the net force and torque on each particle, $(\bR_\Omega^*\tbfv)_i = \left(\int_{\Omega_i}\tbfv_i ~ \d S, \int_{\Omega_i}(\bx-\bx_{\cm,i})^\perp\cdot\tbfv_i ~ \d S\right)$. To solve \eqref{eq:stokes-system}, we need a method to discretize the operators $\bS_\Omega$ and $\grad B_\Gamma$. The primary difficulty is the diagonal term $\bS_{\Omega_i}[\tbfv_i]$, which requires areal integration of a singular kernel against a density with an inverse square-root singularity. 

Our approach uses the projected spherical harmonics (PSH), which are a smooth orthonormal basis on the unit disk under the $L^2$ inner product with weight $1/\omega$ where $\omega(\bxi) = \sqrt{1-|\bxi|^2}$. In the coordinate system $\bxi(\varrho,\vartheta) = \sqrt{1 - \varrho^2}(\cos\vartheta,\sin\vartheta)$, these functions are given by
\begin{equation}
    y^m_\ell(\bxi) = N_\ell^m P^{|m|}_\ell\left(\varrho\right) e^{\i m \vartheta} , \quad 0 \le |m|\le \ell ~ \text{and} ~ \ell + m ~ {\rm even},\label{eq:psh}
\end{equation}
where $P_\ell^{m}$ are the associated Legendre polynomials and $N_\ell^m$ is a normalization constant (equation \eqref{eq:N_lm}). Arising from solutions to the Laplace equation in oblate spheroidal coordinates, the projected spherical harmonics satisfy generalized eigenfunction relations for the Laplace single layer potential, defined over a general set $B\subset\bbR^2$ as
\begin{equation}
    S_B[\nu](\bx) := \int_B G_{\rm 3D}(\bx-\by) \nu(\by) ~ \d S_{\by},\label{eq:laplace-single-layer}
\end{equation}
on the unit disk, where $G_{\rm 3D}(\br) = 1/(4\pi|\br|)$ is the fundamental solution of the 3D Laplace equation. Defining $\alpha(\br) = \arctan(r_2,r_1)$, in the case that $n$ is even, for a point $\bxi \in D$ we have
\begin{equation}
    S_D\left[e^{\i n\alpha(\bxi-\betav)}\frac{y_\ell^m(\betav)}{\omega(\betav)}\right](\bxi) = \lambda_{\ell}^{m,n} y_\ell^{m+n}(\bxi),\label{eq:laplace-single-layer-eig}
\end{equation}
where $\lambda_\ell^{m,n}$ is a constant (equation \eqref{eq:lambda_lmn}) and we use the notation $S_B[\nu(\bx,\by)] = \int_B G_{\rm 3D}(\bx-\by) \nu(\bx,\by) ~ \d S_{\by}$ with $\by$ or $\betav$ the integration variable \citep{wolfe1971eigenfunctions,boersma1993solution}.

The Stokes single layer potential over the reference ellipse $\Omega_0$ may be expressed in terms of $S_D$ as
\begin{equation}
    \begin{aligned}
    \bS_{\Omega_0}[\bnu](\bx_0(\bxi)) &= 
    \int_D \bG(\bx_0(\bxi) - \bx_0(\betav)) \bnu(\bx_0(\betav)) ~ ab\d S_{\betav}
    \\ & =
    \int_D \bA_0(\alpha(\bxi-\betav))G_{\rm 3D}(\bxi-\betav) \bnu(\bx_0(\betav)) ~ \d S_{\betav}
    \\ & =
    S_D[\bA_0(\alpha(\bxi-\betav))\bnu(\bx_0(\betav))](\bxi),
    \end{aligned}\label{eq:stokes-single-layer-decomp}
\end{equation}
where $\bA_0(\alpha) = ab(\bI+\hat\bgamma_0(\alpha)\hat\bgamma_0(\alpha))/|\bgamma_0(\alpha)|$. This form is suitable for the spectral relation \eqref{eq:laplace-single-layer-eig} and only requires the Fourier coefficients of $\bA_0(\alpha)$, which are nonzero only for even modes since $\bA_0(\alpha+\pi)=\bA_0(\alpha)$ so that \eqref{eq:laplace-single-layer-eig} indeed applies. The operator over $\Omega_i$ has the same structure with $\bA_i(\alpha) = \bQ(\theta_{\cm,i})\bA_0(\alpha)\bQ(\theta_{\cm,i})^T$. For each $\tbfv_i$, we thus make the expansion $\tbfv_i(\bx_i(\bxi_i)) = (1/\omega(\bxi_i)) \sum_{\ell+m~{\rm even}} \widehat{\bfv}_{i,\ell}^m y_\ell^m(\bxi_i)$ for $\bxi_i \in D$, which encodes the inverse square-root singularity and can be computed efficiently using orthogonality of the PSH. Combining this expansion with the decomposition \eqref{eq:stokes-single-layer-decomp} and the spectral relation \eqref{eq:laplace-single-layer-eig}, the Stokes single layer potential can then be evaluated rapidly with high accuracy. This spectral formulation applies for the diagonal blocks; off-diagonal blocks are smooth on the interior and are evaluated with the same expansion for $\tbfv_i$ and the quadrature rule \eqref{eq:psh-quadrature}.

When forming the right hand side of the integral equation, we need the PSH expansion $\grad B_\Gamma[\sigma](\bxi_i) = \sum_{\ell,m} \widehat{\grad B}{}_\ell^m y_\ell^m(\bxi_i)$ on each $\Omega_i$. However, $\grad B_\Gamma[\sigma]$ is non-smooth up to $\Gamma$ so that computing the coefficients directly by the quadrature rule \eqref{eq:psh-quadrature} is inaccurate. Instead, we exchange the order of integration, converting the singular integral into an integral on $\Gamma$,
\begin{equation}
  \widehat{\grad B}{}_\ell^m = \int_\Gamma \sigma(\by) \bh_\ell^m(\by) ~ \d\Gamma_{\by},\label{eq:gradB-hat}
\end{equation}
where
\begin{equation}
  \bh_\ell^m(\bx) = \frac{1}{ab}\left(S_\Omega\left[\by  y_\ell^{m*}/\omega\right](\bx) - \bx  S_\Omega\left[y_\ell^{m*}/\omega\right](\bx)\right)
\end{equation}
is smooth. (Here $S_\Omega[\by y_\ell^{m*}/\omega]$ is evaluated by computing the PSH expansion of $\by(\betav) y_\ell^{m*}(\betav)$ using the orthogonality relation.) The remaining integrand in equation \eqref{eq:gradB-hat} is therefore of the same regularity as $\sigma$, and can be computed with the quadrature used to solve for $\sigma$ (see below). 

\subsection{Representation of the surfactant density}

Defining the restrictions $\sigma_D = \sigma\rvert_{\Gamma_D}$ and $\sigma_N = \sigma\rvert_{\Gamma_N}$, the surfactant problem can similarly be written as the block system
\begin{equation}
\begin{pmatrix}
    \cV_{\Gamma_D} & \cV_{\Gamma_N} & I \\
    \cK'_{\Gamma_N,\Gamma_D} & \cK'_{\Gamma_N} & 0 \\
    I^*_{\Gamma_D} & I^*_{\Gamma_N} & 0
\end{pmatrix}
\begin{pmatrix}
\sigma_D \\ \sigma_N \\ c_\infty
\end{pmatrix}
=
\begin{pmatrix} g \\ 0 \\ 0 \end{pmatrix},\label{eq:surfactant-system}
\end{equation}
where $\sigma_D = (\sigma_{1,D},\ldots,\sigma_{N,D})$, $\sigma_N = (\sigma_{1,N},\ldots,\sigma_{N,N})$, and $g = (g_1,\ldots, g_N)$, and each layer potential is an $N\times N$ block operator whose entries are the corresponding potentials from particle $j$ evaluated on particle $i$. Explicitly, $(\cV_{\Gamma_D})_{ij}[\sigma_{j,D}] = \cV_{\Gamma_{j,D}}[\sigma_{j,D}]\rvert_{\Gamma_{i,D}}$ for $i\neq j$ and $(\cV_{\Gamma_D})_{ii} = V_{\Gamma_{i,D}}$. Similarly $(\cK'_{\Gamma_N})_{ij} = \cK'_{\Gamma_{i,N},\Gamma_{j,N}}$ for $i\neq j$ and $(\cK'_{\Gamma_N})_{ii} = -1/2 + K'_{\Gamma_{i,N}}$. Other terms are defined similarly. The operator $I$ adds the constant $c_\infty$ on every Dirichlet arc, and its adjoint $I^*_\Gamma\sigma = \int_\Gamma \sigma ~ \d \Gamma$ enforces the zero net flux condition over all particles. To solve the surfactant system \eqref{eq:surfactant-system}, we need a method to discretize the operators $\cV_{\Gamma_D}$, $\cV_{\Gamma_N}$, $\cK'_{\Gamma_N,\Gamma_D}$ and $\cK'_{\Gamma_N}$. The primary difficulties are in the diagonal terms, which require contour integration of singular kernels against densities with half-integer, possibly singular, powers. 

Our approach uses a change of variables that resolves the singular behavior near the Dirichlet-Neumann junction. To this end, for the arc parameterized by $s \in (s_a,s_b)$, we make the change of variables
\begin{equation}
    s(\varphi) = s_a + \frac{(s_b-s_a)(1 + \cos\varphi)}{2},
\end{equation}
with $\varphi \in (0,\pi)$. Since $s(\varphi) - s_a = (s_b-s_a)\cos^2(\varphi/2)$ and $s_b - s(\varphi) = (s_b-s_a)\sin^2(\varphi/2)$, every half-integer power of the distance to either endpoint is a smooth function of $\varphi$. The density is therefore a smooth function of $\varphi$ divided by $\sin\varphi$, which is canceled by the Jacobian $((s_b-s_a)/2)|\bgamma'(s)|\sin\varphi$, so that $\sigma_D(s(\varphi))|\bgamma'(s(\varphi))|\sin\varphi$ and $\sigma_N(s(\varphi))|\bgamma'(s(\varphi))|\sin\varphi$ are both smooth. We can thus solve for these weighted $\sigma_D$ and $\sigma_N$ directly using methods for smooth densities (appendix \ref{app:laplace}).

\subsection{Implementation}\label{sec:multiple}

We solve the boundary integral equations in two stages. First, we solve for $\sigma$, which is independent of the particle velocities, by discretizing the linear system \eqref{eq:surfactant-system} with Fej\'er nodes on each arc $\Gamma_{i,D}, \Gamma_{i,N}$. The system is inverted directly using a dense solver (backslash in Julia). We then substitute this solution into the Stokes system, which is solved iteratively using GMRES, preconditioned by the diagonal blocks. Close evaluation of the Stokes operator is handled by extending the PSH expansion off of the ellipse (appendix \ref{app:psh-exterior}). Evaluation of the off-diagonal terms in the Stokes system is accelerated with the Fast Multipole Method (FMM) as implemented in the \texttt{FMM3D} library \citep{fmm3d}. In all cases we use a PSH expansion of degree $M = 16$ for the force densities and $N_t = 64$ Fej\'er points on each arc. Time integration is done with the explicit midpoint method, with time step $\Delta t = \max\left(10^{-3}, \min\left(5, \min_{i\neq j} \text{dist}(\Gamma_i,\Gamma_j)/2v_{ij}\right)\right)$, where $v_{ij} = |\bu_{\cm,i}-\bu_{\cm,j}| + \max(a,b)(|\omega_{\cm,i}| + |\omega_{\cm,j}|)$ bounds the relative speed of their boundary points. The upper bound $\Delta t \leq 5$ is large, but is chosen based on the single particle propulsion speed (see below) so that a particle does not move more than one body length per time step. A step is rejected and repeated with half the time step if any gap falls below half its value at the start of the step. The simulation is terminated when $\min_{i\neq j}\text{dist}(\Gamma_i,\Gamma_j) < 10^{-2}$. A refinement study verifying spectral convergence is provided in appendix \ref{app:convergence}.

\section{Propulsion of a single particle}\label{sec:single}

In this section we use the boundary integral formulation to compute the motion of a single Marangoni surfer in terms of its boundary surfactant concentration. Propulsion speeds have been calculated analytically for the disk by directly solving the partial differential equations using Hankel transforms \citep{lauga2012viscous} and for ellipsoidal particles using the reciprocal theorem \citep{masoud2014reciprocal}. Here we compute solutions for elliptical particles directly from the boundary integral equations.

\subsection{Resistance formulation}\label{sec:resistance}

Let $\Omega, \Gamma$ be the domain and boundary of a single particle, not necessarily elliptical, and let $\bmu_1,\bmu_2,\bmu_3$ be auxiliary solutions such that $\bS_\Omega[\bmu_1] = \be_x$, $\bS_\Omega[\bmu_2] = \be_y$ and $\bS_\Omega[\bmu_3] = \bx^\perp$. Multiplying equation \eqref{eq:BIE_1} by $\bmu_i$ and integrating gives
\begin{equation}
\begin{pmatrix} 
\langle \bmu_1, \be_x \rangle & \langle \bmu_1, \be_y \rangle & \langle \bmu_1, \bx^\perp \rangle \\
\langle \bmu_2, \be_x \rangle & \langle \bmu_2, \be_y \rangle & \langle \bmu_2, \bx^\perp \rangle \\
\langle \bmu_3, \be_x \rangle & \langle \bmu_3, \be_y \rangle & \langle \bmu_3, \bx^\perp \rangle \\
\end{pmatrix}
\begin{pmatrix}
u_{\cm,1} \\ u_{\cm,2} \\ \omega_\cm
\end{pmatrix}
=
\begin{pmatrix}
\int_\Gamma h_1 \sigma ~ \d\Gamma \\
\int_\Gamma h_2 \sigma ~ \d\Gamma \\
\int_\Gamma h_3 \sigma ~ \d\Gamma 
\end{pmatrix},
\end{equation}
where $\langle \bu,\bv \rangle = \int_\Omega \bu\cdot\bv ~ \d S$ is the $L^2$ inner product on $\Omega$ and
\begin{align}
    h_i(\by) &= \frac{1}{4\pi} \int_\Omega \bmu_i(\bx) \cdot \frac{\bx - \by}{|\bx - \by|} ~ \d S_{\bx}.
\end{align}
The matrix on the left hand side is the resistance matrix while the right hand side incorporates nonlocal forcing induced by the Marangoni stress. (Note that this force is not simply the local Marangoni force $\int_\Gamma c \hat\bn ~ \d\Gamma$, which generally over-predicts the propulsion speed as noted by \cite{lauga2012viscous}.) This formulation is closely related to the reciprocal formulation of \cite{masoud2014reciprocal}, which, for a symmetric particle translating along the $\be_x$ axis, may be expressed as
\begin{equation}
\bu_{\cm} = -\frac{1}{\langle \bmu_1, \be_x \rangle} \left(\int_\Gamma c \hat\bn\cdot\be_x ~ \d\Gamma + \int_{S\setminus\ol{\Omega}} \bS_\Omega[\bmu_1] \cdot \grad c ~ \d S\right)\be_x.
\end{equation}
The advantage of the present formulation is the reduction of the second integral, which is of infinite support, to an integral supported on $\Gamma$.

\subsection{Elliptical particles}

Now consider the case that $\Omega = \Omega_0$, $\Gamma = \Gamma_0$ is the ellipse centered at the origin with major axis half length $a$ and minor axis half length $b \le a$; a similar argument applies for $a < b$. Define the weight $\omega(\bx) = \sqrt{1 - x_1^2/a^2 - x_2^2/b^2}$ and eccentricity $\epsilon = \sqrt{1 - (b/a)^2}$. We assume the boundary surfactant concentration is symmetric about the orientation axis so that $\omega_\cm = 0$. From equation \eqref{eq:Stokes-eig-decomp}, we find $\bmu_i = \hat\mu_i(\be_i/\omega)$ for some constant $\hat\mu_i$. Then $\langle \bmu_i, \be_i \rangle = \hat\mu_i \left(\int_\Omega 1/\omega ~ \d S\right) = 2\pi a b\hat\mu_i$. Moreover, using equation \eqref{eq:laplace-single-layer-eig}, we find
\begin{equation}
\begin{aligned}
    h_i(\bx) &= \hat\mu_i\left(S_\Omega\left[\frac{x_i}{\omega}\right] - x_iS_\Omega\left[\frac{1}{\omega}\right]\right)
    \\ & =
    \hat\mu_i k_i x_i,
\end{aligned}
\end{equation}
where $k_1 = b/(2\epsilon^2)\left[(1-\epsilon^2)\mathsf{K}(\epsilon) - \mathsf{E}(\epsilon)\right]$ and $k_2 = b/(2\epsilon^2)\left[\mathsf{E}(\epsilon) - \mathsf{K}(\epsilon)\right]$ with $\mathsf{K}$ and $\mathsf{E}$ the complete elliptic integrals of the first and second kind, respectively. The center of mass velocity is then given by
\begin{equation}
u_{\cm,i} = \frac{k_i}{2\pi a b} \left(\int_\Gamma x_i \sigma ~ \d\Gamma\right).\label{eq:ucm-ellipse}
\end{equation}

\subsection{Exact solution}\label{sec:exact}

For the surfactant profile $g(s) = -\cos s$, \cite{lauga2012viscous} found
\begin{equation}
    \bu_{\cm} = \frac{\pi}{32}\left(1 + \cos s_0\right)^2 \be_x,\label{eq:ucm-disk}
\end{equation}
where $s_0$ is the intersection point of the Dirichlet and Neumann boundaries (in our case $s_0=\pi/2$). This expression follows immediately from equation \eqref{eq:ucm-ellipse} by noting that $k_1 = -\pi/8$ as $\epsilon\to0$ and $-\sigma/2 = \lim_{r\to 1^+}\partial c/\partial r$. 

We can build on this solution to determine the propulsion speed analytically for an ellipse of arbitrary aspect ratio. Solutions for flat elliptical particles are accessible through the reciprocal approach of \cite{masoud2014reciprocal}, though not considered in that work. Here, the solution arises naturally in the boundary integral formulation by relating the far-field behavior to the multipole moments associated with propulsion. We first introduce the elliptic coordinate system $(\xi, s)$ such that $\bx(\xi,s) = {a\epsilon}(\cosh\xi\cos s, \sinh\xi\sin s)$ with the particle boundary given by the level set $\xi = \xi_0 = {\rm arccosh}(1/\epsilon)$. This coordinate system is conformal so that $c$ satisfies $(\partial_\xi^2 + \partial_s^2)c = 0$ exterior to the particle, $c(\xi_0,s) = g(s)$ on the Dirichlet boundary, and $\partial c/\partial \xi = 0$ on the Neumann boundary. The concentration is therefore given by
\begin{equation}
    c(\xi, s) = c_{\rm disk}(e^{\xi - \xi_0},s)
\end{equation}
where $c_{\rm disk}(r,\theta)$ is the solution exterior to the unit disk in polar coordinates.

Now the propulsion speed depends only on the moment $\bmv_1 := \int_\Gamma \by \sigma(\by) ~ \d\Gamma_{\by}$, which can be determined from the far-field behavior of the solution. Expanding the integral representation \eqref{eq:c=V[sigma]} for large $|\bx|$ and using $\int_\Gamma \sigma ~ \d\Gamma = 0$ gives
\begin{equation}
    c(\bx) = c_\infty + \frac{1}{2\pi}\frac{(\bx\cdot\bmv_1)}{|\bx|^2} + O(|\bx|^{-2}).
\end{equation}
By symmetry $\bmv_1\cdot\be_y = 0$ so that $c \sim c_\infty + (m_1\cos\theta)/(2\pi|\bx|)$ with $m_1 = \bmv_1\cdot\be_x$ and $\theta$ the polar angle. As $\xi \to \infty$, we have $|\bx| = (a\epsilon/2)e^{\xi}\left(1 + O(e^{-2\xi})\right)$ and $\theta = s + O(e^{-2\xi})$. Since $a\epsilon e^{\xi_0} =a\epsilon(\cosh\xi_0 + \sinh\xi_0) = a + b$, this gives $|\bx| \sim [(a+b)/2]e^{\xi-\xi_0}$. Substituting this asymptotic expression into the far-field expansions of $c$ and $c_{\rm disk}$ and matching shows $m_1 = [(a+b)/2] m_{1,\rm disk}$. Finally, substituting this and the disk solution \eqref{eq:ucm-disk} into equation \eqref{eq:ucm-ellipse} gives the velocity of the ellipse,
\begin{equation}
    \bu_\cm = -\frac{(a+b)k_1}{8ab}\left(1 + \cos s_0\right)^2 \be_x.\label{eq:ucm-exact}
\end{equation}
This procedure works for arbitrary boundary surfactant concentrations, allowing the determination of the propulsion speed of the ellipse from the corresponding disk solution.

\section{Far-field approximation}\label{sec:far-field}

The integral formulation shows the surface flow due to the Marangoni effect is captured by the operator $\grad B_\Gamma$. This compressed formulation naturally leads to a multipole expansion of the flow generated by a single particle, which we derive in this section. This expansion also leads to a reduced model for the particle center of mass dynamics.

\subsection{Multipole expansion}\label{sec:multipole}

Because $\int_{\Omega_i} \tbfv_i ~ \d S = \bzero$, the leading order contribution from $\bS_{\Omega_i}[\tbfv_i]$ is a dipole and hence decays as $O(r^{-2})$, where $r$ is the separation distance. On the other hand, the Marangoni forcing $\grad B_\Gamma[\sigma]$ is {\em a priori} $O(1)$. Expanding the kernel $(\bx-\by)/|\bx-\by|$ for $|\bx|\gg|\by|$ gives the approximation to the far-field flow,
\begin{equation}
\begin{aligned}
\bu_i^\infty(\br) &= \frac{1}{4\pi}\left(\int_{\Gamma_i}\sigma_i ~ \d\Gamma\right)\hat\br - \left(\frac{\bI-\hat\br\hat\br}{4\pi r}\right)\cdot\left(\int_{\Gamma_i} (\by-\bx_{\cm,i}) \sigma_i ~ \d\Gamma\right) + O(r^{-2})
\\ & =:
\frac{\hat\br}{4\pi}m_{0,i}  - \left(\frac{\bI-\hat\br\hat\br}{4\pi r}\right)\cdot \bmv_{1,i} + O(r^{-2}).
\end{aligned}
\label{eq:u-far-field}
\end{equation}
The first term generates a radial flow, uniform in magnitude, with strength $m_{0,i} = \int_{\Gamma_i} \sigma_i ~ \d\Gamma$ and is thus related to the surfactant flux. This coefficient is zero for a single particle, but will generally be nonzero for interacting particles. The second term is related to self-propulsion via equation \eqref{eq:ucm-ellipse}, and generates a flow that decays as $1/r$. The induced flow notably produces no disturbance along the propulsive axis of the particle.

\begin{figure}
    \includegraphics[width=\linewidth]{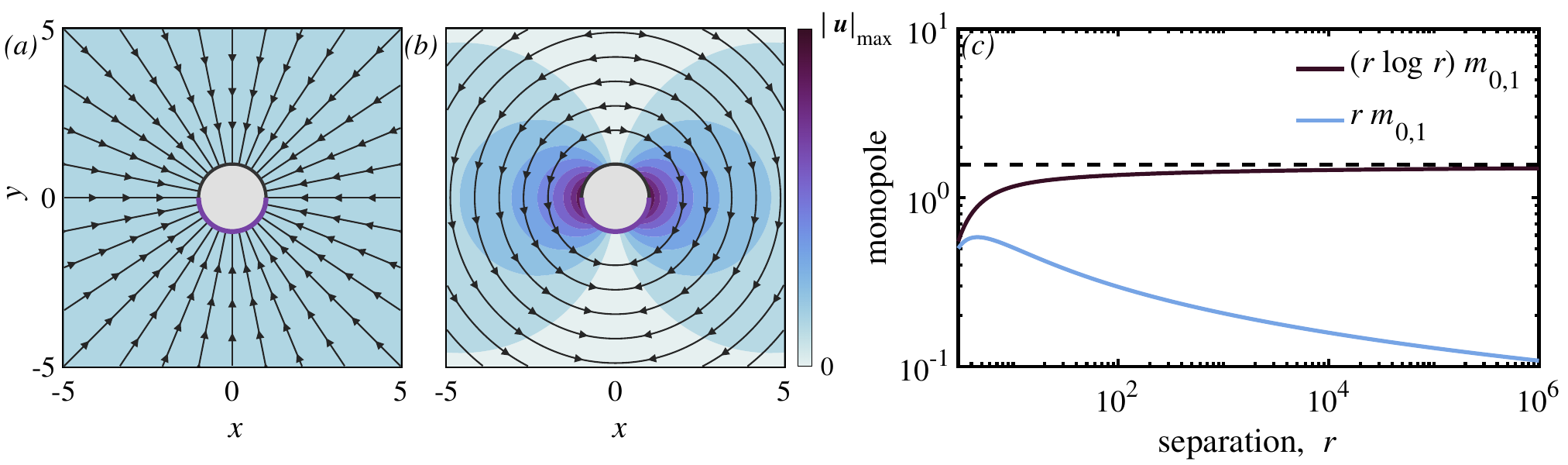}
    \caption{Multipole expansion. {\em(a)} Zeroth and {\em(b)} first order terms in the multipole expansion. Streamlines for the former are radial while streamlines for the latter are circular. {\em(c)} Coefficient $m_{0,1} = \int_{\Gamma_1} \sigma ~ \d \Gamma$ as a function of separation distance $r$ for a pair of particles located at $\bx_{\cm,1} = (-r/2,0)$ and $\bx_{\cm,2} = (r/2,0)$ and orientation $\theta_{\cm,i} = 0$. We find $m_{0,1} = O(1/r\log r)$. The dashed black line indicates the asymptote $\pi/2$.}\label{fig:multipole}
\end{figure}

Figure \ref{fig:multipole}{\em(a)}-{\em(b)} shows the flow fields associated with the first two terms in the multipole expansion. The flow generated by the first term, as discussed above, is purely radial. The second term, on the other hand, generates circular streamlines whose direction is mirrored about the propulsion axis of the particle. Note the flow is nonetheless irrotational due to the gradient structure of $\grad B_\Gamma$. The asymptotic behavior of the multipole coefficient $m_{0,i}$, which is generally nonzero for interacting particles, is characterized in figure \ref{fig:multipole}({\em c}). We show the monopole coefficient under two rescalings, $rm_{0,1}(r)$ and $r\log r m_{0,1}(r)$, computed numerically for a pair of particles of shared orientation $\theta_{\cm,i} = 0$ and center of mass separation $r$. The former tends to zero at large $r$ while the latter approaches $\pi/2$ (black dashed line), suggesting $m_{0,i} = O(1/r\log r)$, subdominant to the $O(1/r)$ dipolar term.

We can understand this scaling directly from the boundary integral equations. For particles with the same shape and boundary conditions, the self-interaction term is identical in the particle frame. In the case of two particles, subtracting the respective boundary integral equations and expanding the cross interaction for $r \gg 1$ gives
\begin{align}
V_{\Gamma_0}[\delta\sigma] + \frac{\log r}{2\pi}\langle \delta\sigma \rangle_{\Gamma_0} = O(1/r) &\quad {\rm on} ~ \Gamma_{0,D},\\
-\frac{\delta\sigma}{2} + K'_{\Gamma_0}[\delta\sigma] = O(1/r) &\quad{\rm on} ~ \Gamma_{0,N},
\end{align}
where $\delta\sigma = \sigma_1 - \sigma_2$, $\Gamma_0$ is the reference boundary in the particle frame, and $\langle \cdot \rangle_{\Gamma_0} = \int_{\Gamma_0}\cdot~\d\Gamma$. Now let $\tau$ solve the single particle problem $V_{\Gamma_0}[\tau] = 1$ on $\Gamma_{0,D}$ and $-\tau/2 + K'_{\Gamma_0}[\tau] = 0$ on $\Gamma_{0,N}$. We then find $\delta\sigma = \delta\sigma' - (\langle \delta\sigma \rangle_{\Gamma_0} \log r/2\pi) \tau$ for some $\delta\sigma' = O(1/r)$. Integrating over $\Gamma_0$ gives
\begin{equation}
  \langle \delta\sigma \rangle_{\Gamma_0} \left[1 + \frac{\log r}{2\pi}\left(\int_{\Gamma_0}\tau ~ \d\Gamma\right)\right] = \langle \delta\sigma' \rangle_{\Gamma_0} = O(1/r),
\end{equation}
which implies $\langle \delta\sigma \rangle_{\Gamma_0} = O(1/r\log r)$ so long as $\langle \tau \rangle_{\Gamma_0} \neq 0$. Since $\langle \sigma_1 \rangle_{\Gamma_0} = -\langle \sigma_2 \rangle_{\Gamma_0}$, we have $m_{0,1} = \langle \delta\sigma \rangle_{\Gamma_0}/2 = O(1/r\log r)$, consistent with numerical observation. This same reasoning shows the overall correction in $\sigma_i$ with respect to the single particle solution is also $O(1/r\log r)$.

\subsection{Far-field interactions}

For well-separated elliptical particles with surfactant symmetrically distributed about their symmetry axis $\bp(\theta_{\cm,i}) = (\cos\theta_{\cm,i},\sin\theta_{\cm,i})$, we found the multipole coefficients $\int_{\Gamma_i} \sigma_i ~ \d\Gamma = O(1/r\log r)$ and $\int_{\Gamma_i} (\by-\bx_{\cm,i}) \sigma_i ~ \d\Gamma = -\eta \bp(\theta_{\cm,i}) + O(1/r\log r)$ for some explicitly computable $\eta > 0$ depending on the boundary surfactant concentration. The expansion can in principle be truncated here, however the $O(1/r\log r)$ corrections can in fact be determined explicitly from the boundary integral equations. Writing $\bmv_{1,j} = -\eta_j\bp_j$ with $\bp_j = \bp(\theta_{\cm,j})$ and expanding the potential due to particle $j$ about $\bx_{\cm,i}$ gives
\begin{equation}
\cV_{\Gamma_j}[\sigma_j](\bx) = -\frac{m_{0,j}}{2\pi}\log|\br_{ij}| - \frac{\eta_j}{2\pi}\frac{\hat\br_{ij}\cdot\bp_j}{|\br_{ij}|} + O(r_{ij}^{-2}), \quad \bx\in\Gamma_i,
\end{equation}
where $\br_{ij} = \bx_{\cm,i} - \bx_{\cm,j}$. To $O(r_{ij}^{-2})$, the cross interaction is constant on $\Gamma_i$, and the normal derivative is $O(r_{ij}^{-2})$. The density on particle $i$ is therefore $\sigma_i = \sigma_0 + \alpha_i\tau$, where $\sigma_0$ is the single particle density, $\tau$ is the response to a uniform shift in the far-field from section \ref{sec:multipole}, and, to $O(r_{ij}^{-2})$,
\begin{equation}
\alpha_i = -\left[c_\infty' + \sum_{j\neq i}\cV_{\Gamma_j}[\sigma_j](\bx_{\cm,i})\right],\label{eq:alpha}
\end{equation}
with $c_\infty'$ the change in $c_\infty$ relative to an isolated particle. Because all particles are identical, we may define $\int_{\Gamma_i} \tau ~ \d\Gamma = \tau_0$ and, by symmetry, $\int_{\Gamma_i}(\by - \bx_{\cm,i})\tau~\d\Gamma = \tau_1\bp(\theta_{\cm,i})$. For elliptical particles, the coefficients can be computed explicitly using elliptical coordinates (section \ref{sec:exact}) and matching the far-field, giving $\tau_0 = 2\pi/\log[4/(a+b)]$ and $\tau_1 = -[(a+b)/2]\tau_0$. The multipole coefficients are therefore $m_{0,i} = \tau_0\alpha_i$ and $\eta_i = \eta - \tau_1\alpha_i$. It remains to determine the $\alpha_i$. Substituting the far-field expansion of $\cV_{\Gamma_j}[\sigma_j]$ and these multipole coefficients into \eqref{eq:alpha} gives a linear system for $\alpha_i$ and $c_\infty'$,
\begin{align}
\alpha_i + \sum_{j\neq i}\left(-\frac{\tau_0}{2\pi}\log|\br_{ij}| + \frac{\tau_1}{2\pi}\frac{\hat\br_{ij}\cdot\bp_j}{|\br_{ij}|}\right)\alpha_j + c_\infty' &= \frac{\eta}{2\pi}\sum_{j\neq i}\frac{\hat\br_{ij}\cdot\bp_j}{|\br_{ij}|},\label{eq:closure}\\
\sum_{i=1}^N\alpha_i &= 0,\label{eq:closure-mass}
\end{align}
where the second equation follows from \eqref{eq:BIE_N}.

Using these multipole coefficients, we can view the center of mass dynamics for particle $i$ as that of a force-free surfer in the background flow \eqref{eq:u-far-field} generated by all other particles. Linearizing the far-field flow about $\bx_{\cm,i}$ and using the result of \cite{jeffery1922motion} gives an approximation for the translational and angular velocities, accurate to $O(r^{-2})$,
\begin{align}
\bu_{\cm,i} &\approx U\frac{\eta_i}{\eta}\bp_i + \sum_{j\neq i}\left[\frac{m_{0,j}}{4\pi}\hat\br_{ij} + \left(\frac{\bI-\hat\br_{ij}\hat\br_{ij}}{4\pi |\br_{ij}|}\right)\cdot \eta_j\bp_j\right],\label{eq:ucm-far-field}\\
\omega_{\cm,i} &\approx \sum_{j\neq i} \beta\bp^\perp_i \cdot\left[\frac{m_{0,j}(\bI - \hat\br_{ij}\hat\br_{ij})}{4\pi|\br_{ij}|} - \frac{\eta_j\left[(\bI - 3\hat\br_{ij}\hat\br_{ij})(\hat\br_{ij}\cdot\bp_j) + \hat\br_{ij}\bp_j + \bp_j\hat\br_{ij}\right]}{4\pi |\br_{ij}|^2}\right]\cdot\bp_i,\label{eq:omegacm-far-field}
\end{align}
where $U$ is the single particle propulsion speed and $\beta = (a^2 - b^2)/(a^2 + b^2)$ is a geometric factor arising from the resistance of an ellipsoidal particle with one principal axis taken to zero. The self-propulsion speed is rescaled by $\eta_i/\eta$ since, by \eqref{eq:ucm-ellipse}, the self-induced velocity depends only on the particle's own dipole moment. These expressions, together with the linear system \eqref{eq:closure}--\eqref{eq:closure-mass}, define a system of $3N$ ordinary differential equations (ODEs) for the positions and orientations of $N$ particles.

\section{Pair interactions}\label{sec:pair}

In this section we perform direct numerical simulations of pairs of surfers. We use the multipole expansion to rationalize our observations, which simultaneously serves to check the range of applicability of the far-field model.

\begin{figure}
    \includegraphics[width=\linewidth]{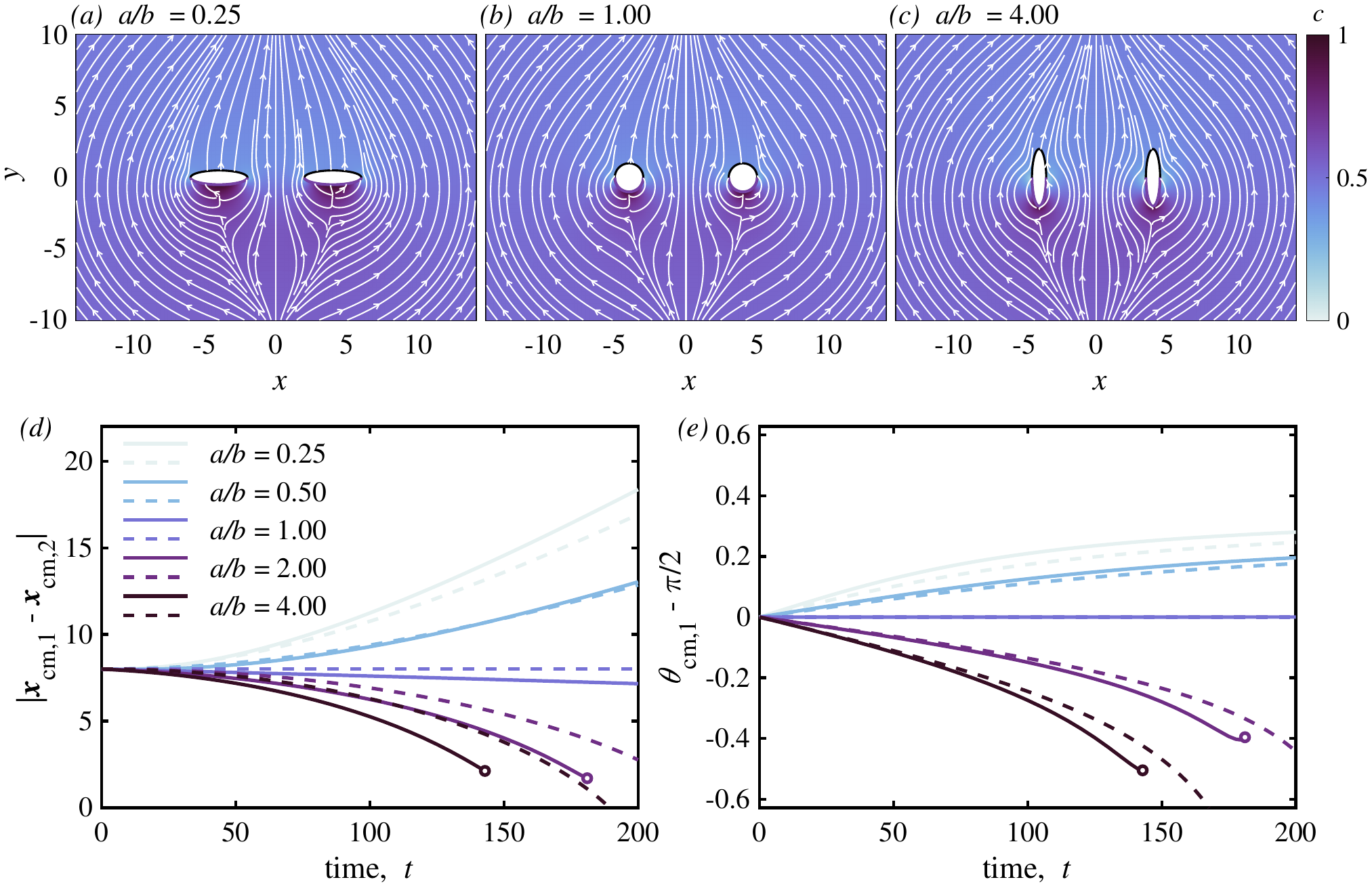}
    \caption{Dynamics of parallel surfers. {\em(a)}-{\em(c)} Flow and surfactant fields for pairs of parallel surfers with separation $d_0=8$ and aspect ratios $a/b = 0.25, 1$ and $4$, respectively. The flow and surfactant fields are qualitatively similar in each case, though streamlines undergo less deflection in the prolate case. {\em(d)} Separation distance for particles with various aspect ratios initially separated by $d_0 = 8$ from direct numerical simulation (solid) and the far-field model (dashed). Prolate particles $(a>b)$ attract while oblate particles $(a<b)$ repel. On the same timescales, the disks $(a=b)$ do not significantly change their separation distance. {\em(e)} Approach at higher aspect ratios occurs due to rotation.}\label{fig:pair_parallel}
\end{figure}

\subsection{Parallel surfers}

We first consider pairs initially positioned at $\bx_{\cm,1} = (-d_0/2,0)$ and $\bx_{\cm,2} = (+d_0/2,0)$ and oriented in a parallel configuration $\theta_{\cm, 1} = \theta_{\cm, 2} = \pi/2$. Figure \ref{fig:pair_parallel} shows the surface flow and surfactant fields generated by such pairs with $d_0=8$ for aspect ratios $a/b = 0.25, 1$, and $4$. The flow is qualitatively similar in each case, with slight variability in the deflection of the streamlines near the body. All cases tend towards circular streamlines in the far-field, consistent with the multipole expansion. The surfactant distribution is also similar across each case, with gradients, and consequently Marangoni stresses, localized near the particle boundaries.

The temporal dynamics of parallel pairs with initial separation distance $d_0 = 8$ are characterized in figure \ref{fig:pair_parallel}{\em(d)}-{\em(e)}. Panel {\em(d)} shows the center of mass separation distance over time (solid lines). As the particles self-propel, prolate particles $(a>b)$ move towards one another, seemingly colliding in finite time (open circle), while oblate particles $(a<b)$ move away from one another. The disks $(a=b)$ do not significantly change their separation distance on these timescales. These behaviors are due to pair-induced rotation, clarified by the evolution of the reference angle $\theta_{\cm,i}$ shown in panel {\em(e)} (solid lines). Higher aspect ratio particles rotate faster, with prolate particles rotating towards one another and oblate particles rotating away from one another. The sense of these rotations is in agreement with the far-field model \eqref{eq:omegacm-far-field}, which predicts $\dot\theta_{\cm,1} \propto -\beta$ for the leftmost particle, with $\beta > 0$ for prolate particles, $\beta < 0$ for oblate particles, and $\beta = 0$ for disks. We compare the direct numerical simulations with simulations of the far-field model (dashed lines). The far-field model captures both the translational and orientational dynamics for well-separated particles, but as expected departs from the direct numerical simulation as particles approach.

\subsection{Anti-parallel surfers}

\begin{figure}
    \includegraphics[width=\linewidth]{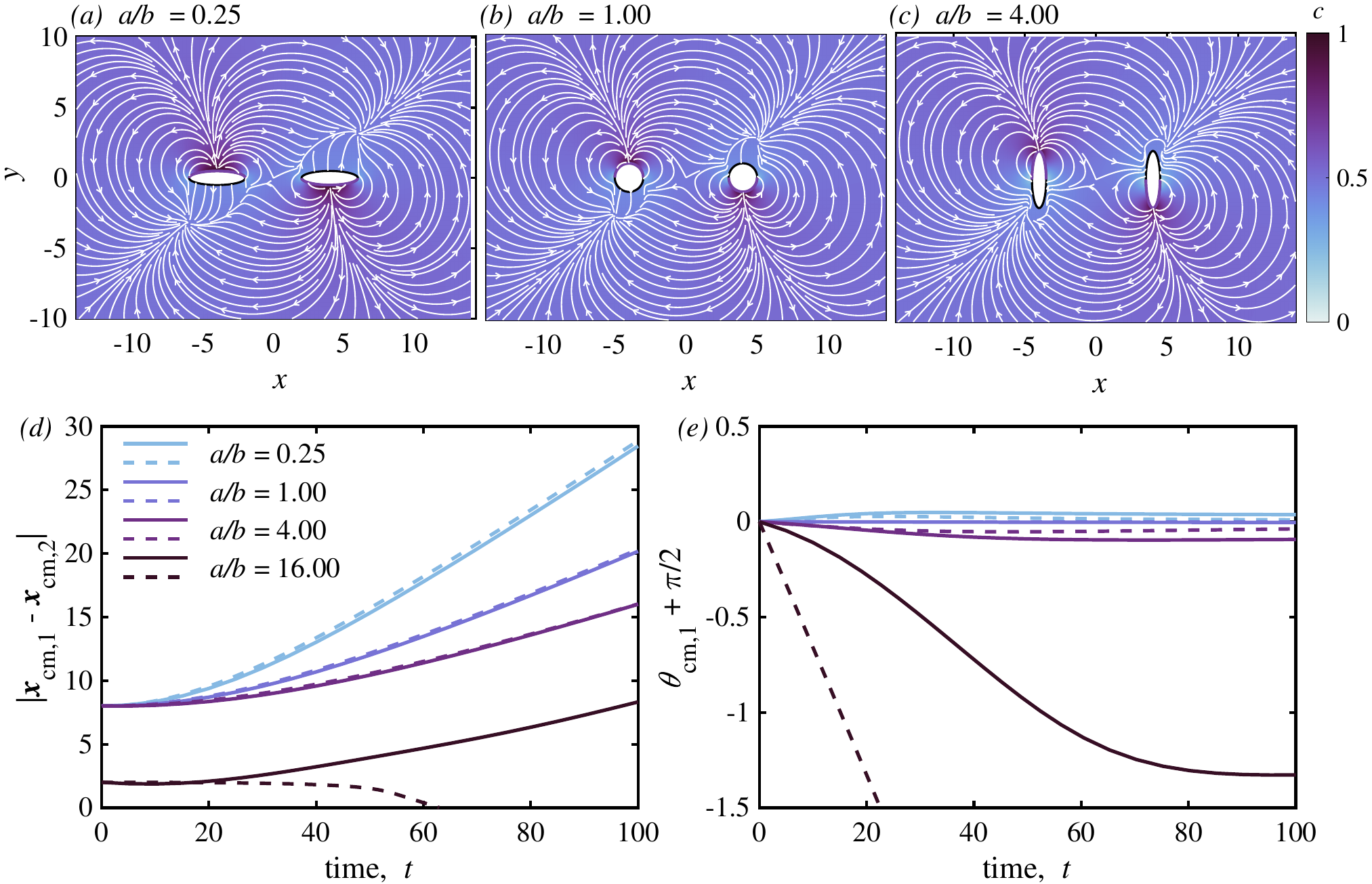}
    \caption{Dynamics of anti-parallel surfers. {\em(a)}-{\em(c)} Flow and surfactant fields for pairs of anti-parallel surfers with separation distance $d_0=8$ and aspect ratios $a/b = 0.25, 1$ and $4$, respectively. All cases show a four-fold structure, and surfactant gradients are localized near the particle boundaries. {\em(d)} Separation distance for particles with various aspect ratios and initial separation distances from direct numerical simulation (solid) and the far-field model (dashed). {\em(e)} Prolate particles undergo initial rotation before drifting apart, while oblate and disk-shaped particles separate without significant rotation.}\label{fig:pair_antiparallel}
\end{figure}

We next consider pairs initially positioned at $\bx_{\cm,1} = (-d_0/2,0)$ and $\bx_{\cm,2} = (+d_0/2,0)$, here oriented in an anti-parallel configuration $\theta_{\cm, 1} = -\pi/2$ and $\theta_{\cm, 2} = \pi/2$. Figure \ref{fig:pair_antiparallel}{\em(a)}-{\em(c)} shows the surface flow and surfactant fields for $d_0 = 8$. In contrast to the parallel configuration, the flow exhibits a four-fold structure, though the detailed structures of the streamlines differ near the particle boundaries. 

We similarly show the temporal dynamics in panels {\em(d)} and {\em(e)} for particles with aspect ratios $a/b = 0.25, 1, 4$ for the same initial separation as above, as well as $a/b = 16$ at a closer distance. In each case particles separate, and the rate of separation decreases with the aspect ratio. This is more clearly illustrated by the angular dynamics in panel {\em(e)}, which for the pair with $a/b = 16$, shows the particles initially rotate about one another before drifting apart. For the simulations that remain well-separated, the far-field model provides a good approximation, particularly in the translational dynamics.

\begin{figure}
    \centering
    \includegraphics[width=0.5\linewidth]{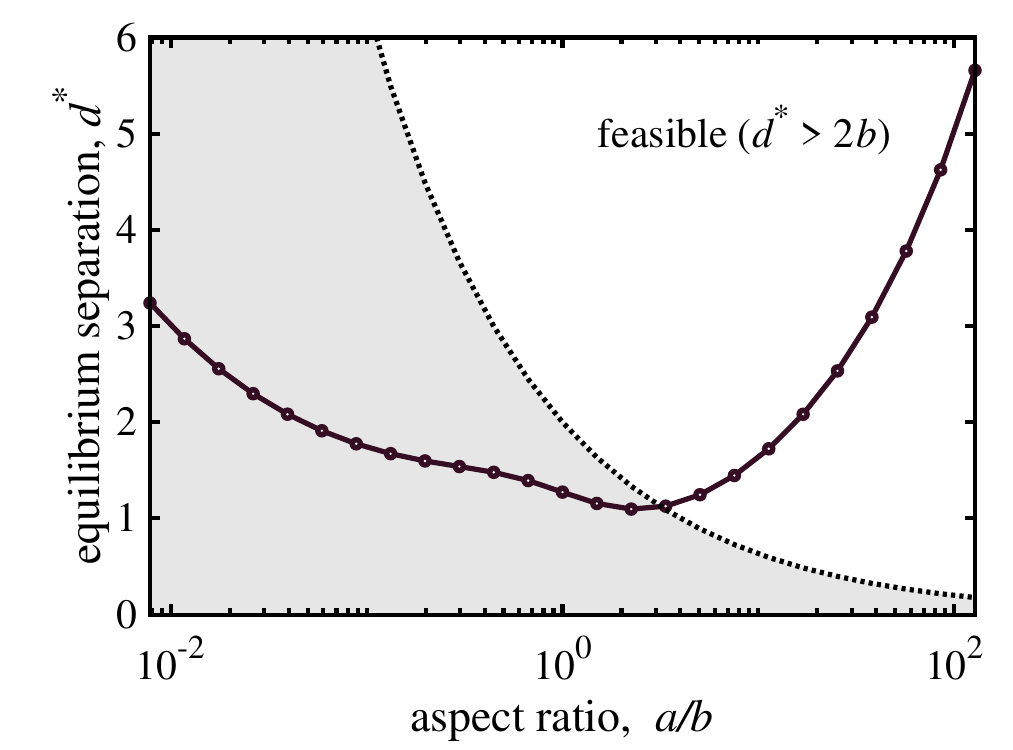}
    \caption{Equilibrium distance for an orbiting pair of particles as a function of the aspect ratio based on the far-field approximation. The gray region corresponds to a distance smaller than twice the particle width and is therefore unphysical. We find physically realizable solutions are only achievable for prolate particles.}\label{fig:equilibrium_separation}
\end{figure}

We can gain further insight into this behavior by analyzing the far-field ODEs \eqref{eq:ucm-far-field}--\eqref{eq:omegacm-far-field}. Making the Ansatz $\br_{12} = -d^*\bp_1^\perp$ for some constant $d^*$ and $\theta_{\cm,2} = \theta_{\cm,1} + \pi$, and using $\dot\br_{12} = \bu_{\cm,1} - \bu_{\cm,2}$ and $\dot\bp_i^\perp = -\omega_{\cm,i}\bp_i$, we find an exact solution
\begin{align}
d^* &= \frac{\eta}{4\pi U} \left(1 - \frac{\beta}{2}\right),\\
\br_{12}(t) &= d^*\begin{pmatrix} \sin(\theta_{\cm,1}(t)) \\ -\cos(\theta_{\cm,1}(t)) \end{pmatrix},\\
\theta_{\cm,1}(t) &= -\frac{\beta\eta}{4\pi (d^*)^2} t + \theta_{\cm, 1}(0),\\
\theta_{\cm,2}(t) &= -\frac{\beta\eta}{4\pi (d^*)^2} t + \theta_{\cm, 2}(0).
\end{align}
This corresponds to a pair of particles orbiting at a fixed separation distance $d^*$, but is only realizable when $d^* > 2b$. (Note that, for elliptical particles, equation \eqref{eq:ucm-ellipse} shows the ratio $\eta/U$ is independent of the surfactant density $\sigma$ and hence $d^*$ is determined by geometry alone.) For disks, $\beta = 0$, this solution predicts the particles remain stationary. On the other hand, prolate ellipses, $\beta > 0$, orbit in the clockwise direction while oblate ellipses, $\beta < 0$, rotate in the counter-clockwise direction. The range of feasibility as a function of the particle aspect ratio, with $\eta$ and $U$ computed from the single-particle problem, is shown in figure \ref{fig:equilibrium_separation}. We find only prolate particles, $a/b\gtrapprox3$, lie in the feasible range, and only extreme aspect ratios, $a/b\gg 1$, have equilibrium separations at which the multipole expansion may apply, though these are inaccessible through our current numerical implementation. Nonetheless, the qualitative character of this solution agrees with the early time behavior observed in figure \ref{fig:pair_antiparallel}.

\section{Collective interactions}\label{sec:collective}

In this final section, we consider collective interactions among a large number of particles. We first perform direct numerical simulations with the boundary integral formulation, again comparing our observations to the far-field predictions.

\subsection{Expansion of a circular patch}

\begin{figure}
    \includegraphics[width=\linewidth]{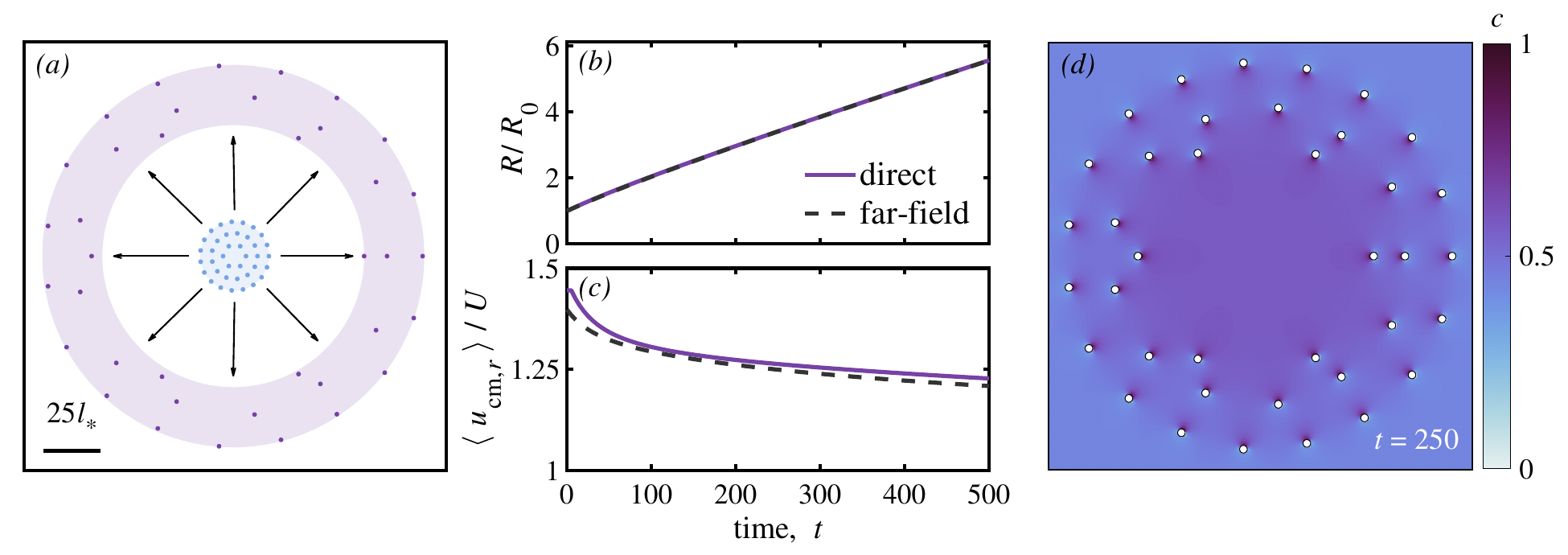}
    \caption{Expansion of a concentrated collection of disks. {\em(a)} Snapshots of the configuration at $t = 0$ and $t = 500$. {\em(b)} The maximal radius of the configuration increases approximately linearly in time. {\em(c)} The average radial velocity is larger than the single particle propulsion speed and gradually decreases in time. {\em(d)} Snapshot of the surfactant field at time $t = 250$ shows a system-scale gradient in the surfactant concentration, which drives a global Marangoni flow.}\label{fig:expansion}
\end{figure}

Experiments of {\em B. subtilis} populating a free surface show surfactant production can enhance expansion and wall climbing, even in the absence of flagellar motility \citep{angelini2009bacillus}. Motivated by these observations, we simulate a collection of disks, here $N = 38$, initially positioned in a circular patch and oriented radially outward. Random initial orientations produce similar dynamics but typically result in close interactions that are challenging to resolve numerically. Figure \ref{fig:expansion}({\em a}) shows snapshots of the configuration at $t = 0$ and $t = 500$. The areal extent of the population, whose structure remains roughly axisymmetric over time, is indicated by the shaded annular region. The maximal radius of the configuration, shown in panel {\em(b)}, increases approximately linearly in time, consistent with ballistic motion. Strikingly, the average radial velocity, shown in panel {\em(c)}, is nearly 50\% larger than the single particle propulsion speed, and gradually decreases as the population expands. The mechanism behind this speed increase can be discerned from the surfactant field, shown in panel {\em(d)} for the configuration at time $t = 250$. On the interior of the patch, the surfactant concentration is relatively uniform and, importantly, higher than outside of the patch. This produces system scale surfactant gradients, and hence Marangoni stresses, that result in a net radially outward flow.

We also compare this expansion to the far-field model, initialized with the same configuration. The radial growth (panel {\em(b)}) is nearly identical to the direct simulation, growing linearly in time. The average radial velocity (panel {\em(c)}) is also within a few percent of the direct simulation, though slightly under-predicts the speed up. 

\subsection{Stability of circular patches}

\begin{figure}
    \includegraphics[width=\linewidth]{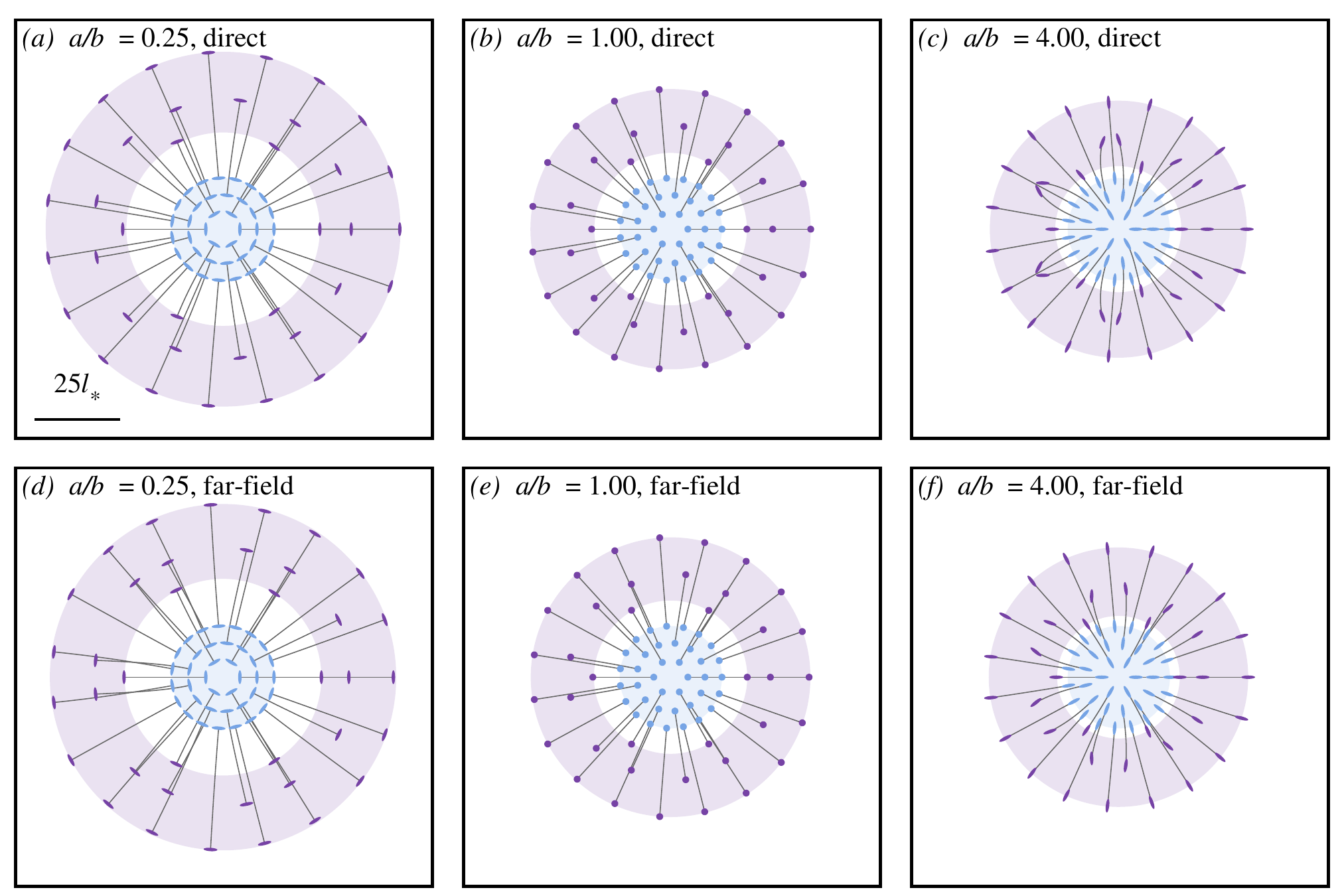}
    \caption{Expansion of an initially circular patch for {\em(a)} oblate particles ($a/b = 0.25$), {\em(b)} disks ($a/b = 1$), and {\em(c)} prolate particles ($a/b = 4$). Blue corresponds to the initial configuration while purple corresponds to the configuration at $t = 180$, near the apparent collision time of the prolate case. Solid lines are the individual particle trajectories. The oblate and disk-shaped particles expand stably while prolate particles aggregate. Panels {\em(d)}-{\em(f)} show results of the far-field approximation for the same conditions as {\em(a)}-{\em(c)}, which are in close agreement.}\label{fig:expansion_aspect}
\end{figure}

The stable radial expansion of the patch of disk-shaped particles is suggested from the far-field approximation, which predicts the angular velocity vanishes for well-separated disks. Figure \ref{fig:expansion_aspect} assesses the evolution of the same circular configuration for oblate particles ($a/b = 0.25$) and prolate particles ($a/b = 4$) in addition to the disk. In the former case, shown in panel {\em(a)}, we find the patch expands stably with little reorientation as demonstrated by the radial trajectories (solid black lines). The disk case, shown in panel {\em(b)}, exhibits similarly straight trajectories though expands at a slightly slower rate. In contrast, prolate particles, shown in panel {\em(c)}, exhibit interior reorientations that ultimately lead to aggregation and apparent collisions within the patch. The far-field model, panels {\em(d)}-{\em(f)}, shows nearly identical behavior, though over-predicts the apparent collision time of the prolate case. These behaviors are suggested by the pair interactions, where we found aligned prolate particles rotated towards each other while oblate particles rotated away from each other.

\subsection{Stability of tandem chains}

\begin{figure}
    \includegraphics[width=\linewidth]{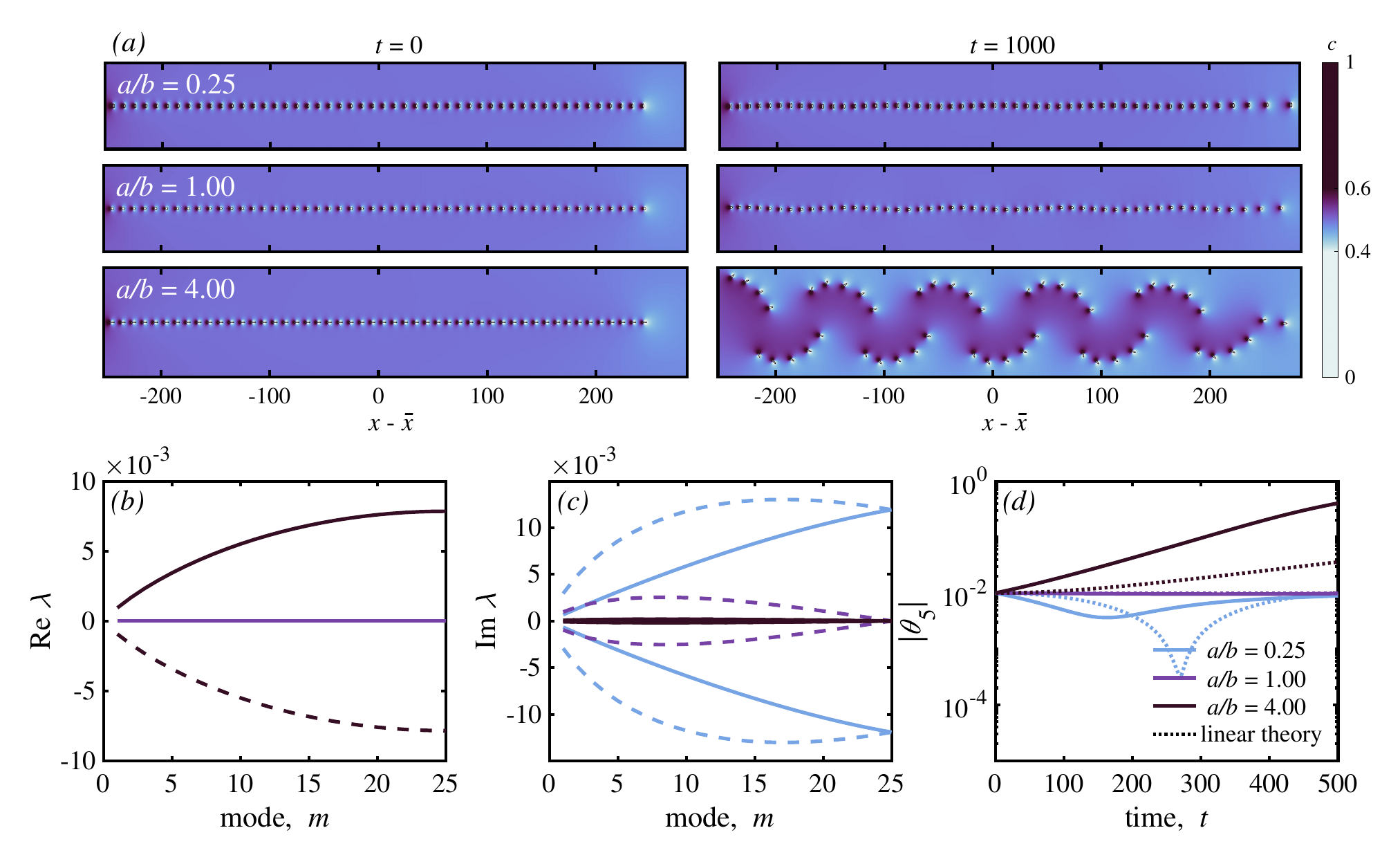}
    \caption{Dynamics of tandem surfers. {\em(a)} Snapshots of the particle configuration and surfactant field (color) for $a/b=0.25, 1, 4$, relative to the center of mass of the configuration. The perturbation grows only in the prolate case ($a/b=4$). The {\em(b)} real and {\em(c)} imaginary parts of the dispersion relation of the far-field model capture the qualitative behavior of the direct particle simulations. {\em(d)} Evolution of the $m = 5$ mode shows qualitative agreement between the particle simulations (solid) and far-field linear theory (dotted) with difference in timescales.}\label{fig:tandem}
\end{figure}

The circular patch simulations show particle aspect ratio is closely tied to the stability of many-particle systems. To probe this more directly, we consider a chain of tandem particles initially positioned and oriented along the $x$-axis, $\bx_{\cm,i} = (id,0)$ and $\bp_i = \be_x$ for $i = 1,\ldots,N$, with inter-particle spacing $d$. Figure \ref{fig:tandem} shows snapshots of a chain of $N = 50$ particles separated by $d = 10$ to a periodic orientation perturbation of $m = 5$ wavelengths with amplitude 0.01 at times $t = 0$ and $t = 1000$. For oblate particles (top row), the orientation angle fluctuates slightly but the particles roughly remain in a line. Disk-shaped particles (middle row) behave similarly, with no discernible reorientation. For prolate particles (bottom row), in sharp contrast, the perturbation grows and the colony destabilizes at the wavelength of the perturbation, breaking up into smaller chains of particles that themselves break up.

We can understand these differences through a stability analysis of the far-field model for an infinite chain. We consider a perturbation about a translating chain $\bx_{\cm,j}(t) = (jd + Ut,{\rm Re}(\hat y e^{\lambda t + \i \tk j}))$ and $\theta_{\cm,j}(t) = {\rm Re}(\hat\theta e^{\lambda t + \i \tk j})$ with $j = \ldots, -1, 0, 1, \ldots$ and $\tk = kd = 2\pi m/N \in (0,\pi]$, where $k = 2\pi m/(Nd)$ is the wavenumber of mode $m$ and $\hat y, \hat\theta \ll 1$. Substituting this expression into the center of mass dynamics \eqref{eq:ucm-far-field}--\eqref{eq:omegacm-far-field}, we get an eigenvalue problem for the growth rate $\lambda$,
\begin{equation}
    \begin{pmatrix}
      -\dfrac{\i\eta}{2\pi d^2}\displaystyle\sum_{j=1}^\infty\frac{\sin j\tk}{j^2} & U + \dfrac{\eta}{2\pi d}\displaystyle\sum_{j=1}^\infty\frac{\cos j\tk}{j} \\
      \dfrac{\beta\eta}{\pi d^3}\displaystyle\sum_{j=1}^\infty\frac{1-\cos j\tk}{j^3} & \dfrac{\i\beta\eta}{2\pi d^2}\displaystyle\sum_{j=1}^\infty\frac{\sin j\tk}{j^2}
    \end{pmatrix}
    \begin{pmatrix}\hat y\\ \hat\theta\end{pmatrix}
    =
    \lambda\begin{pmatrix}\hat y\\ \hat\theta\end{pmatrix},
  \end{equation}
which can be evaluated numerically. Note the matrix is complex because the chain is not symmetric under the perturbation, so $\cos j\tk$ modes couple to $\sin j\tk$ modes, producing traveling waves. The two eigenvalues $\lambda_\pm$ are therefore not complex conjugates, but rather the matrix at $-\tk$ is the complex conjugate of that at $\tk$. Thus a real perturbation of mode $m$, which combines $\pm\tk$, has the four eigenvalues $\lambda_\pm$ and $\lambda_\pm^*$. Figure \ref{fig:tandem}{\em(b)}-{\em(c)} shows the real and imaginary parts, respectively, as a function of the mode number for aspect ratios $a/b=0.25,1,4$. We find oblate and disk-shaped particles have purely imaginary eigenvalues, while the prolate case has a complex conjugate pair of eigenvalues with positive real part. We compare this linear theory (dotted lines) to the direct numerical simulations in panel {\em(d)}, which shows the amplitude of the $m = 5$ mode over time. Without fitting, the linear theory qualitatively captures an initial reorientation of the oblate case, the unchanged disk case, and the instability of the prolate case, though the time scales differ.

\section{Discussion}\label{sec:discussion}

We developed a boundary integral formulation for the analysis and simulation of Marangoni surfers in the low Reynolds and P\'eclet number limit. This formulation generalizes established results on viscous Marangoni propulsion \citep{lauga2012viscous,masoud2014reciprocal} to flat elliptical particles, and has the benefit of extending single particle analytical and computational methods to many-particle systems. The particle-supported formulation in particular lends itself to a multipole expansion whose form is difficult to discern otherwise owing to the infinite extent of the Marangoni stress. This expansion captures pair and collective interactions observed in direct particle simulations, and provides a tractable coarse-grained model in the large particle number limit.

For an initially circular patch of radially aligned surfers, we found, as in two dimensions \citep{crowdy2020collective}, collective interactions enhance the rate of expansion. Moreover, radial expansion was stable for oblate and disk-shaped particles while prolate particles aggregated. While microbial systems, such as the colonies of {\em B. subtilis} considered by \cite{angelini2009bacillus}, typically consist of prolate particles for which our analysis predicts an instability, steric interactions at high number densities may stabilize expansion. Incorporating steric effects is an interesting direction for future work, and would require a model for post-contact dynamics.

Our boundary integral simulations consist of a moderate number of particles, but do not reach the numbers or densities required for the chaotic dynamics typical of active suspensions to occur \citep{saintillan2013active}. Whether such dynamics persist in this system is unclear, though the central ingredients -- flow-induced alignment, self-propulsion, and long-range interactions -- are all present. In this large particle number limit, the far-field approximation can be used to derive a continuum kinetic theory. Such theories can be used to not only simulate many-particle systems, but also analyze their stability. Developing and analyzing such a model is the subject of ongoing work.

The spectral treatment of the Stokes operator generalizes to other bulk-surface fluid structure interaction problems. This includes discrete models of particles with prescribed kinematics, such as pusher or puller particles confined to a single layer, but also continuum models of active chiral droplets, lipid domains, or growing microbial communities \citep{stone1995hydrodynamics,lubensky1996hydrodynamics,jia2022incompressible,herrera2026instability}. In these latter contexts, departure from the elliptical geometry will require additional techniques, such as conformal mapping, in order to exploit the spectral properties of the disk operator. 

Finally, a fundamental limitation of the boundary integral formulation is its restriction to the vanishing Reynolds and P\'eclet number limit. This limit is a plausible approximation for microbial systems, but is likely inappropriate for larger particles such as aquatic insects or engineered surfers. Nonetheless, this formulation could be exploited in small $\Rey$ and $\Pec$ expansions as an efficient base state solver or preconditioner.

\begin{bmhead}[Acknowledgements]
We thank Vicente Gomez Herrera for useful discussions. We used Claude Opus (Anthropic) to implement and optimize portions of the code, assist with and check analytical calculations, and proofread the manuscript. All text was written by the author. The computations in this work were performed at facilities supported by the Scientific Computing Core at the Flatiron Institute, a division of the Simons Foundation.
\end{bmhead}

\begin{bmhead}[Data Availability]
    The data that support the findings of this study are openly available at \url{https://github.com/scottweady/marangoni-surfers.git}.
\end{bmhead}

\begin{bmhead}[Declaration of Interests]
    The author reports no conflict of interest.
\end{bmhead}

\begin{appen}

\section{Bulk diffusion}\label{app:bulk-diffusion}

Here we outline an analogous boundary integral formulation when the chemical species diffuses in the bulk.  Let $C : V \to \bbR$ be the concentration of the species. For small $\Pec$ number, here with respect to the bulk diffusion constant, $C$ satisfies the Laplace equation $-\lapv C = 0$ with boundary conditions
\begin{align}
C = g &\quad {\rm on} ~ \Omega,\\
\frac{\partial C}{\partial z} = 0 &\quad {\rm on} ~ S\setminus\ol{\Omega},\\
C \to 0 &\quad {\rm as} ~ |\bX|\to\infty.
\end{align}
We write $C(\bX) = \int_\Omega G_{\rm 3D}(\bX - (\by,0))\sigma(\by) ~ \d S_{\by}$. This representation satisfies $-\lapv C = 0$ in the bulk, $\partial C/\partial z = 0$ on $S\setminus\ol{\Omega}$, and $C\to 0$ as $|\bX|\to\infty$. The density $\sigma$ can then be determined by solving the integral equation $S_\Omega[\sigma] = g$ on $\Omega$.

As before, we use the integral representation of the velocity field, which on the surface is given by $\bu = \bS_\Omega[\bfv] - \bS_{S\setminus\ol{\Omega}}[\grads C]$, and define the modified force $\tbfv = \bfv + \grads C$ so that $\bu = \bS_\Omega[\tbfv] - \bS_S[\grads C]$. Using $\cF[G_{\rm 3D}\rvert_{z=0}] = 1/(2k)$, we compute
\begin{equation}
 \grads C = \cF^{-1}\left[\i\bk \frac{\widehat{\chi_\Omega \sigma}}{2k}\right].   
\end{equation}
Further, using $\cF[\bG] = \bI/k -\bk\bk/(2k^3)$, we find
\begin{equation}
    \cF[\bG * \grads C] = \frac{\i\bk}{4k^2}\cF[\chi_\Omega \sigma].
\end{equation}
Taking an inverse transform, the surface velocity is given by
\begin{equation}
    \bu(\bx) = \bS_\Omega[\tbfv](\bx) + \grads\left(\frac{1}{8\pi}\int_\Omega \log|\bx-\by| \sigma(\by) ~ \d S_{\by}\right).
\end{equation}
The main difference between this formulation and the surface diffusion formulation is the $\log|\bx|$ kernel in contrast to the $|\bx|$ kernel. This results in the leading order far-field behavior $\propto\hat\br/r \left(\int_\Omega \sigma ~ \d S\right)$ compared to $\propto(\bI-\hat\br\hat\br)/r\cdot\left(\int_\Gamma \by\sigma ~ \d\Gamma\right)$. Note this former scaling was identified by \cite{wurger2014thermally} using the explicit solution for a sphere.

\section{Sublimation}\label{app:sublimation}

Surfactant sublimation, at low P\'eclet number, can be described by the modified Helmholtz equation \citep{crowdy2021viscous},
\begin{equation}
-\laps c + \kappa^2 c = 0 \quad{\rm on} ~ S \setminus \ol{\Omega},\label{eq:helmholtz}
\end{equation}
subject to the same boundary conditions as the pure diffusion case, where $\kappa$ is a rate constant. We express $c$ as a Helmholtz single layer potential
\begin{equation}
c(\bx) = \frac{1}{2\pi} \int_\Gamma K_0(\kappa|\bx-\by|) \sigma(\by) ~ \d\Gamma_{\by},
\end{equation}
where $K_0$ is the modified Bessel function of the second kind. This automatically satisfies \eqref{eq:helmholtz}. Following the same approach as in the pure Laplace case, we find
\begin{equation}
\cF[\bG * \grads\tc] = \frac{\i\bk}{2k(k^2 + \kappa^2)}\cF[\delta_\Gamma \sigma].
\end{equation}
The inverse transform of $1/[k(k^2 + \kappa^2)]$ is
\begin{equation}
\cF^{-1}\left[\frac{1}{k(k^2 + \kappa^2)}\right] = \frac{1}{4\kappa} \left[ I_0(\kappa|\bx|) - L_0(\kappa|\bx|) \right],
\end{equation}
where $I_0$ is the modified Bessel function of the first kind and $L_0$ is the modified Struve function. The surface velocity is then given by 
\begin{equation}
\bu(\bx) = \bS_\Omega[\tbfv](\bx) - \grads\left( \frac{1}{8\kappa}\int_\Gamma \left[ I_0(\kappa|\bx-\by|) - L_0(\kappa|\bx-\by|) \right] \sigma(\by) ~ \d\Gamma_{\by} \right).
\end{equation}
This screened version differs from the pure diffusion case in that the kernel decays as $O(1/r)$ rather than $O(r)$ in the far-field, hence its contribution to the flow is $O(1/r^2)$ rather than $O(1)$.

\section{Bulk flow}\label{app:bulk-flow}

After taking a Fourier transform in the plane, the Stokes equations reduce to a system of ordinary differential equations in $z$. The solution can be computed analytically, and is given by \citep{herrera2026instability}
\begin{align}
\hat\bu(\bk,z) &= \left[ \hat\bu_0 - \frac{\i\bk}{k}(\i\bk\cdot\hat\bu_0)z\right] e^{kz},\\
\hat w(\bk, z) &= -(\i\bk\cdot\hat\bu_0) z e^{kz},\\
\hat p(\bk, z) &= -2(\i\bk\cdot\hat\bu_0) e^{kz},
\end{align}
where
\begin{equation}
\hat\bu_0(\bk) = \frac{1}{k}\left(\bI - \frac{\bk\bk}{2k^2}\right)\cdot(\widehat{\chi_\Omega\tbfv}) - \frac{\i\bk}{2k^3}\widehat{(\delta_\Gamma\sigma)} =: \hat\bu_0^S + \hat\bu_0^M,
\end{equation}
is the Fourier transform of the surface velocity. The inverse transform of the first term is the standard convolution with the Stokeslet. For the second term, we compute $\i\bk\cdot\hat\bu_0^M = \widehat{(\delta_\Gamma\sigma)}/(2k)$ so that
\begin{align}
\hat\bu^M(\bk, z) &= -\i\bk\left(\frac{1}{2k^3} + \frac{z}{2k^2}\right)\widehat{(\delta_\Gamma\sigma)}e^{kz}, \\
\hat w^M(\bk, z) &= -\left(\frac{z}{2k^2}\right)\widehat{(\delta_\Gamma\sigma)}e^{kz}.
\end{align}
The real space velocity is then
\begin{equation}
\bU = \bG_{\rm 3D}*\begin{pmatrix} 2\chi_\Omega\tbfv \\ 0 \end{pmatrix} - \begin{pmatrix} \grads \left[\left( \frac{1}{2}\Phi_{-3} + \frac{z}{2}\Phi_{-2} \right) * (\delta_\Gamma\sigma)\right] \\
 \frac{z}{2} \Phi_{-2} * (\delta_\Gamma\sigma)
\end{pmatrix},
\end{equation}
where $\Phi_n(r,z) = \cF^{-1}[k^n e^{kz}]$. The inverse transforms for $\Phi_{-3}$ and $\Phi_{-2}$ are not classically defined and must be interpreted in the sense of distributions. We first compute the classical inverse transform $\Phi_{-1}(r, z) = 1/(2\pi \sqrt{r^2 + z^2})$. Using $-\laps \Phi_{-3} = \Phi_{-1}$ and solving in polar coordinates, we find
\begin{align}
    2\pi \Phi_{-3}(r, z) &= -\sqrt{r^2 + z^2} - z\log\left(\sqrt{r^2 + z^2} - z\right),\\
    2\pi \Phi_{-2}(r, z) &= -\log\left(\sqrt{r^2 + z^2} - z\right),
\end{align}
up to constants of integration, where we defined $\Phi_{-2} = \partial_z \Phi_{-3}$. Because $\int_\Gamma \sigma ~ \d\Gamma = 0$, the constants of integration do not contribute and can be set to zero without loss of generality. Note this recovers the surface velocity in the main text in the limit $z\to 0^-$.

\section{Stokes layer potentials}\label{app:stokes}

In this section we describe further details of the numerical method used to evaluate and invert the Stokes single layer potential.

\subsection{Projected spherical harmonics}

First recall the projected spherical harmonics in the coordinate system $\bxi(\varrho,\vartheta) = \sqrt{1 - \varrho^2}(\cos\vartheta,\sin\vartheta)$ \eqref{eq:psh},
\begin{equation}
    y_\ell^m(\bxi) = N_\ell^m P^{|m|}_\ell\left(\varrho\right) e^{\i m \vartheta} , \quad 0 \le |m|\le \ell ~ \text{and} ~ \ell + m ~ {\rm even}.
\end{equation}
Under the normalization constant
\begin{equation}
    N_\ell^m = \sqrt{ \frac{2\ell+1}{2\pi} \frac{(\ell - |m|)!}{(\ell + |m|)!} }, \label{eq:N_lm}
\end{equation}
these functions are orthonormal under the inner product
\begin{equation}
\langle y_\ell^m, y_{\ell'}^{m'} \rangle_\omega = \int_D \frac{y_\ell^{m}(\bx) y_{\ell'}^{m'*}(\bx)}{\omega(\bx)} ~ \d S.\label{eq:psh-orthogonality}
\end{equation}
The generalized eigenvalues of the single layer potential in this basis are given by
\begin{equation}
\lambda_\ell^{m,n} =
\frac{\varphi_\ell^{m}~\varphi_\ell^{m+n}}{2^{n+2}}
\frac{
\upGamma\!\left(\frac{\ell+m+1}{2}\right)
\upGamma\!\left(\frac{\ell-m-n+1}{2}\right)
}{
\upGamma\!\left(\frac{\ell-m}{2}+1\right)
\upGamma\!\left(\frac{\ell+m+n}{2}+1\right)
}
\sqrt{
\frac{(\ell-m)!(\ell+m+n)!}
{(\ell+m)!(\ell-m-n)!}
}, \quad |m| \le \ell ~\text{and}~ |m+n| \le \ell, \label{eq:lambda_lmn}
\end{equation}
where $\upGamma$ is the Gamma function and
\begin{equation}
\varphi_\ell^k =
\begin{cases}
1 & 0 \le k \le \ell,\\
(-1)^k & -\ell \le k < 0,\\
0 & {\rm otherwise}
\end{cases}
\end{equation}
is a phase factor.

We expand all functions on the disk in the PSH basis, truncating at degree $M$ such that $|m| \le \ell \le M$. The coefficients are computed numerically from the orthogonality relation \eqref{eq:psh-orthogonality}, using Gauss-Legendre quadrature in $\varrho$ and the trapezoidal rule in $\vartheta$. Note that in this coordinate system $\d S/\omega = \d\varrho\d\vartheta$, which removes the inverse square-root singularity. For any smooth $\nu$, the quadrature rule is given explicitly by
\begin{equation}
    \langle \nu, y_\ell^m\rangle_\omega \approx \sum_{i=1}^{N_\varrho} \sum_{j=1}^{N_\vartheta} y_\ell^{m*}(\bxi_{ij})\nu(\bxi_{ij})    w_i w_j,\label{eq:psh-quadrature}
\end{equation}
where $\bxi_{ij} = \bxi(\varrho_i,\vartheta_j)$ with $\varrho_i$ the Gauss-Legendre nodes of order $N_\varrho$, $\vartheta_j$ the equispaced trapezoidal nodes, $w_i$ the Gauss-Legendre weights, and $w_j$ the trapezoidal weights \citep{trefethen2019approximation}. The number of quadrature points is chosen to be $N_\varrho = M+1$ and $N_\vartheta = 2M+1$, which exactly resolves the expansion for polynomials up to degree $M$. In practice, the sum \eqref{eq:psh-quadrature} is computed using the fast Fourier transform in $\vartheta$ and recursion relations for the associated Legendre polynomials in $\varrho$.

\subsection{Spectral representation of the single layer potential}

To evaluate integrals arising in the Stokes single layer potential, we use the decomposition \eqref{eq:stokes-single-layer-decomp} of the integrand together with the spectral relation \eqref{eq:laplace-single-layer-eig}. Writing $\bA_0(\alpha) = \sum_{n} \widehat\bA_n e^{\i n\alpha}$ for the Fourier expansion of $\bA_0$, which may be computed with the fast Fourier Transform, we have, on the reference ellipse $\Omega_0$,
\begin{equation}
    \bS_{\Omega_0}\left[\hat\bnu_\ell^m y_\ell^m / \omega\right](\bx_0(\bxi)) = \sum_n \widehat\bA_n~\lambda_\ell^{m,n}~\hat\bnu_\ell^m~ y_\ell^{m+n}(\bxi).\label{eq:Stokes-eig-decomp}
\end{equation}
Thus, given a weighted expansion of a force density $\tbfv_0(\bx_0(\bxi)) = (1/\omega)\sum_{\ell,m} \hat\bfv_\ell^m y_\ell^m(\bxi)$, the single layer potential can be computed as
\begin{equation}
    \bS_{\Omega_0}[\tbfv_0](\bx_0(\bxi)) = \sum_{\ell,m,n} \widehat\bA_n~\lambda_\ell^{m,n}~\hat\bfv_\ell^m~ y_\ell^{m+n}(\bxi).
\end{equation}
On the other hand, when $\bv =\bS_{\Omega_0}[\tbfv_0]$ is given by the expansion $\bv = \sum_{\ell,m} \hat\bv_\ell^m y_\ell^m$, orthogonality of the $y_\ell^m$ gives the block linear system
\begin{equation}
    \sum_{p}\lambda_\ell^{m-p,p}\widehat{\bA}_p \hat\bfv_\ell^{m-p}
    = 
    \hat\bv_\ell^m, \quad |m| \le \ell, ~ \ell+m ~ {\rm even}.
\end{equation}
This system decouples over $\ell$ and, due to decay of the coefficients $\widehat{\bA}_n$, the sum in $p$ is effectively banded up to a given numerical precision and the linear system can therefore be inverted efficiently.

\subsection{Close evaluation}\label{app:psh-exterior}

The decomposition \eqref{eq:stokes-single-layer-decomp} in fact holds on all of $\bbR^2$. In particular, for a point $\bx \in S$, the preimage $\bxi = \bx_i^{-1}(\bx) = {\rm diag}(1/a,1/b)\,\bQ(\theta_{\cm,i})^T(\bx-\bx_{\cm,i})$ is well-defined and we may write
\begin{equation}
    \bS_{\Omega_i}[\tbfv_i](\bx) = \sum_{\ell,m,n} \widehat\bA_{i,n}~\widehat\bfv_{i,\ell}^m~\psi_\ell^{m,n}(\bxi),\label{eq:stokes-exterior}
\end{equation}
where
\begin{equation}
    \psi_\ell^{m,n}(\bxi) := S_D\left[e^{\i n\alpha(\bxi-\betav)}\frac{y_\ell^m(\betav)}{\omega(\betav)}\right](\bxi), \quad \bxi \in \bbR^2.
\end{equation}
The $\psi_\ell^{m,n}$ can be expressed in terms of hypergeometric functions \citep{olver2010nist}. This expression can be derived using the Fourier transform, similar to the approach of \cite{boersma1993solution}, though we omit the derivation here. Letting $\mu = |m+n|$ and $\chi_k = 1$ for $k \ge 0$, $\chi_k = (-1)^k$ for $k<0$, we have the expression
\begin{equation}
    \psi_\ell^{m,n}(\bxi) = \frac{\sqrt{\pi}}{4}~\chi_{m+n}(-1)^{(\ell-m)/2}N_\ell^m P_\ell^{|m|}(0)~e^{\i(m+n)\vartheta}~
    \upGamma\!\left(\tfrac{\ell+\mu+1}{2}\right) F_\ell^\mu(|\bxi|),\label{eq:psi-closed}
\end{equation}
with
\begin{equation}
    F_\ell^\mu(r) =
    \begin{cases}
    \dfrac{r^\mu}{\upGamma\!\left(\frac{\ell-\mu}{2}+1\right)\upGamma(\mu+1)}~{}_2F_1\!\left(\tfrac{\ell+\mu+1}{2},\tfrac{\mu-\ell}{2};\mu+1;r^2\right), & r < 1,\\[3ex]
    \dfrac{r^{-\ell-1}}{\upGamma\!\left(\frac{\mu-\ell+1}{2}\right)\upGamma\!\left(\ell+\frac{3}{2}\right)}~{}_2F_1\!\left(\tfrac{\ell+\mu+1}{2},\tfrac{\ell-\mu+1}{2};\ell+\tfrac{3}{2};r^{-2}\right), & r > 1.
    \end{cases}
\end{equation}
Equation \eqref{eq:stokes-exterior} is used if the target point satisfies
\begin{equation}
  |\bxi| < 1 + \frac{2\pi c}{2M + 1}\frac{\max(a,b)}{\min(a,b)},
  \label{eq:near-criterion}
\end{equation}
which incorporates both the azimuthal grid spacing $2\pi/(2M+1)$ and the particle aspect ratio $\max(a,b)/\min(a,b)$. At all other targets the standard PSH quadrature \eqref{eq:psh-quadrature} is used. We take $c = 8$, for which the relative error of the PSH quadrature at this distance is below $10^{-8}$ with $M = 16$ for aspect ratios $1 \leq \max(a,b)/\min(a,b) < 10$. The set of near targets depends only on the configuration and is computed once per time step, before the iterative solve. 

\subsection{Spectral representation of the Marangoni operator}\label{app:marangoni-projection}

Recalling that $G_{\rm 3D}(\br) = 1/(4\pi|\br|)$, we can write $\grad B_{\Gamma_0}[\sigma](\bx) = \int_{\Gamma_0} (\bx-\by) G_{\rm 3D}(\bx-\by) \sigma(\by) ~ \d\Gamma_{\by}$. Exchanging the order of integration, the PSH coefficients of $\grad B_{\Gamma_0}[\sigma]$ are
\begin{equation}\begin{aligned}
  \widehat{\grad B}{}_\ell^m &= \int_D \frac{y_\ell^{m*}(\bxi)}{\omega(\bxi)} \grad B_{\Gamma_0}[\sigma](\bx(\bxi)) ~ \d S_{\bxi} \\
  &= \int_{\Gamma_0} \sigma(\by)  \bh_\ell^m(\by) ~\d\Gamma_{\by}.\label{eq:marangoni-projection}
\end{aligned}\end{equation}
Now as in the decomposition of the Stokes operator \eqref{eq:stokes-single-layer-decomp}, we have $S_{\Omega_0}[\nu](\bx_0(\bxi)) = S_D[A_0(\alpha(\bxi-\betav)) \nu(\bx(\betav))](\bxi)$ with $A_0(\alpha) = ab/|\bgamma_0(\alpha)|$. Expanding $A_0(\alpha) = \sum_n \widehat A_n e^{\i n\alpha}$, which can again be computed using the fast Fourier Transform, the eigenfunction relation \eqref{eq:laplace-single-layer-eig} gives
\begin{equation}
  S_{\Omega_0}\left[y_\ell^m/\omega\right](\bx_0(\bxi)) = \sum_n \widehat A_n \lambda_\ell^{m,n} y_\ell^{m+n}(\bxi),
\end{equation}
which is smooth. The remaining integrand in \eqref{eq:marangoni-projection} is therefore as regular as $\sigma$.

\section{Laplace layer potentials}\label{app:laplace}

In this section we describe a method for evaluating the integrals arising in the surfactant integral equation. As in section \ref{sec:numerical-method}, each arc $\Gamma_{i,D}$ and $\Gamma_{i,N}$ is parameterized by $s(\varphi(t)) = s_a + (s_b-s_a)[1 + \cos(\varphi(t))]/2$, with $\varphi(t) = \pi(1-{t})/2$ and ${t}\in(-1,1)$. We define $\tbgamma({t}) = \bgamma(s(\varphi(t)))$ and $\tsigma({t}) = \sigma(s(\varphi(t))) |\tbgamma'({t})|$, so that $\int_\Gamma \nu\sigma ~ \d\Gamma = \int_{-1}^1 \nu\tsigma ~ \d{t}$, the latter of which is as regular as $\nu$. We discretize integrals against smooth functions $\nu$ over the Fej\'er nodes ${t}_j = \cos((2j-1)\pi/(2N_t))$, $j = 1,\dots,N_t$, with corresponding weights $w_j$ \citep{trefethen2019approximation}. This gives the quadrature rule
\begin{equation}
    \int_\Gamma \nu\sigma ~ \d\Gamma \approx \sum_{j=1}^{N_t} \nu({t}_j) \tsigma({t}_j) w_j.
\end{equation}

\subsection{Single layer potential}

For a point on the arc $\tbgamma(t)$, the single layer potential may be expressed as
\begin{equation}
V_\Gamma[\sigma](\tbgamma({t})) = -\frac{1}{2\pi}\int_{-1}^1 \log|\tbgamma({t})-\tbgamma({t}')|  \tsigma({t}') ~ \d {t}'.
\end{equation}
Singularities occur as $s(\varphi(t'))\to s(\varphi(t))$. In terms of $t$, defining $J = (s_b-s_a)/2$, we have
\begin{equation}
  s - s' = 2J\sin\left(\frac{\pi({t}-{t}')}{4}\right)\sin\left(\frac{\pi(2-{t}-{t}')}{4}\right).
\end{equation}
Thus the kernel $\log|\tbgamma(t)-\tbgamma(t')|$ is singular at ${t}' = {t}$ and nearly singular at ${t}' = 2-{t}$ and ${t}' = -2-{t}$. These latter two points lie outside $[-1,1]$ but approach it when ${t}$ is near an endpoint. We therefore decompose the kernel as
\begin{equation}
  \log|\tbgamma({t})-\tbgamma({t}')| = R({t},{t}') + \log|t-t'| + \log|(2-t) - t'| + \log|(-2-t) - t'|,
\end{equation}
where
\begin{equation}
    \begin{aligned}
  R(t,t') &= \log\left(2J\left(a^2\sin^2\bar s + b^2\cos^2\bar s\right)^{1/2}\right) 
  \\ & + \log\left|\frac{\sin((s-s')/2)}{(s-s')/2}\right| + \log\left|\frac{\sin(\pi(t-t')/4)}{t-t'}\right| + \log\left|\frac{\cos(\pi \bar t/2)}{4(1-\bar t^2)}\right|
    \end{aligned}
\end{equation}
is smooth and we used $\bgamma(s)-\bgamma(s') = 2\sin((s-s')/2) (-a\sin\bar s, b\cos\bar s)$ with $\bar s = (s+s')/2$, and $\bar t = (t+t')/2$. This has a removable singularity at $t'\to t$. We thus define
\begin{equation}
  R(t,t) = \lim_{t'\to t} R(t,t') = \log\left(\frac{\pi J}{2}\left(a^2\sin^2 s + b^2\cos^2 s\right)^{1/2}\cos\left(\frac{\pi t}{2}\right)\right) - \log\left(4(1-t^2)\right).
\end{equation}

For the logarithmic terms, we compute a Chebyshev expansion $\tsigma({t}) \approx \sum_{n=0}^{N_t-1} \hat\sigma_n T_n({t})$, where $T_n$ are the Chebyshev polynomials, and express the integrals in terms of the moments
\begin{equation}
  m_n(z) = \int_{-1}^1 \log|z-{t}'| T_n({t}') ~\d {t}',
\end{equation}
where $z$ may be real or complex. The coefficients $\hat\sigma_n$ can be accurately computed using orthogonality of the Chebyshev polynomials, which over the Fej\'er nodes gives $\hat\sigma_n = \sum_{j=1}^{N_t} (\bT^*_{N_t})_{nj} \tsigma({t}_j)$, where $(\bT^*_{N_t})_{nj} = (2-\delta_{n0})T_n({t}_j)/N_t$. This gives the quadrature rule
\begin{equation}
  V_\Gamma[\sigma](\tbgamma({t}_i)) \approx -\frac{1}{2\pi}\sum_{j=1}^{N_t}\left(R({t}_i,{t}_j) w_j + \sum_{n=0}^{N_t-1}\big[m_n({t}_i) + m_n(2-{t}_i) + m_n(-2-{t}_i)\big] (\bT^*_{N_t})_{nj}\right)\tsigma({t}_j),
\end{equation}
which is spectrally accurate for smooth $\tsigma$ so long as $m_n$ is computed spectrally, a method for which we describe in appendix \ref{app:moments}.

\subsection{Adjoint double layer potential}

For $\bx = \bgamma(s)$ and $\by = \bgamma(s')$ on the same ellipse, the identity $\bgamma(s)-\bgamma(s') = 2\sin((s-s')/2) (-a\sin\bar s, b\cos\bar s)$ from above gives
\begin{equation}
  \hat\bn(\bx)\cdot\grad_\bx G(\bx-\by) = -\frac{ab}{4\pi|\bgamma'(s)||\bgamma'(\bar s)|^2},
\end{equation}
which is smooth. Thus both $K'_{\Gamma_N}$ and $\cK'_{\Gamma_N,\Gamma_D}$ can be computed directly with Fej\'er quadrature.

\subsection{Coupling between arcs}

The operator $\cV_{\Gamma_N}$, evaluated on $\Gamma_D$, has a nearly singular kernel near the Dirichlet-Neumann junctions, where the target $\bx$ approaches the end of the source arc. Here we use the approach of \cite{af2021accurate}, which swaps this singularity for one at its complexified parameter preimage that can be handled by the same approach as above. For the ellipse, these values can be determined analytically. Let $s_x$ be the parameter nearest the interval $(s_a,s_b)$ such that $\bx = \bgamma(s_x)$ and define $z_x = 1 - (2/\pi)\arccos\left((s_x - s_a)/J - 1\right)$. The kernel $\log|\bx - \tbgamma({t}')|$ is singular at the complex conjugate pair ${t}' = z_x, \bar z_x$. Subtracting $\log|z_x - {t}'| + \log|\bar z_x - {t}'|$ from the kernel leaves a smooth remainder, and the subtracted terms are integrated exactly with the moments $m_n(z_x)$ and $m_n(\bar z_x)$ defined above.

\subsection{Evaluation of the moments}\label{app:moments}

To evaluate the moments efficiently, we use a recurrence relation analogous to that of \cite{helsing2009integral} for the monomial basis. Let $Q_n(z) = \int_{-1}^z T_n(t') ~ \d t'$ be the antiderivative of $T_n$. Writing $\log|z-{t}'| = {\rm Re}[\log(z-{t}')]$, integration by parts gives
\begin{equation}
  m_n(z) = Q_n(1)\log|1-z| - {\rm Re}\left[Q_n(z)\big(\log(z-1) - \log(z+1)\big) + \frac{1}{2}\left(\frac{D_{n+1}(z)}{n + 1} - \frac{D_{n-1}(z)}{n - 1}\right)\right],\label{eq:m-recurrence}
\end{equation}
where $D_n(z) = \int_{-1}^1 (T_n(z)-T_n(t'))/(z-t') ~ \d t'$, and for $n = 0$ and $n = 1$ the last term is replaced by $D_1$ and $D_2/4$, respectively. The three-term recurrence relation for the $T_n$ gives $D_{n+1} = 2zD_n - D_{n-1} + 2Q_n(1)$, with $D_0 = 0$ and $D_1 = 2$.

The expression \eqref{eq:m-recurrence} holds for all $z$ but is unstable away from the interval $(-1,1)$. To remedy this, define $t' = \cos\theta'$ and $w = z - \sqrt{z^2-1}$, with the branch of the square root chosen so that $|w| < 1$ and $z = (w + w^{-1})/2$. We then have the expansion
\begin{equation}
  \log|z - t'| = -\log|2w| - 2\sum_{k=1}^\infty \frac{{\rm Re}(w^k)}{k}\cos k\theta'.
\end{equation}
Defining $F_n = \int_0^\pi \cos(n\theta')\sin\theta' ~ \d\theta'$, with $F_n = 2/(1-n^2)$ for even $n$ and $F_n = 0$ for odd $n$, we get
\begin{equation}
  m_n(z) = -F_n\log|2w| - \sum_{k=1}^\infty \frac{{\rm Re}(w^k)}{k}\left(F_{k+n} + F_{k-n}\right).
\end{equation}
The series converges geometrically and requires at most $O(N_t)$ terms in the regime where it is used. In practice, we find the recursion formula \eqref{eq:m-recurrence} is accurate when $|w|^{-N_t} < e^5$, otherwise the series is used.

\section{Convergence}\label{app:convergence}

\begin{figure}
    \includegraphics[width=\linewidth]{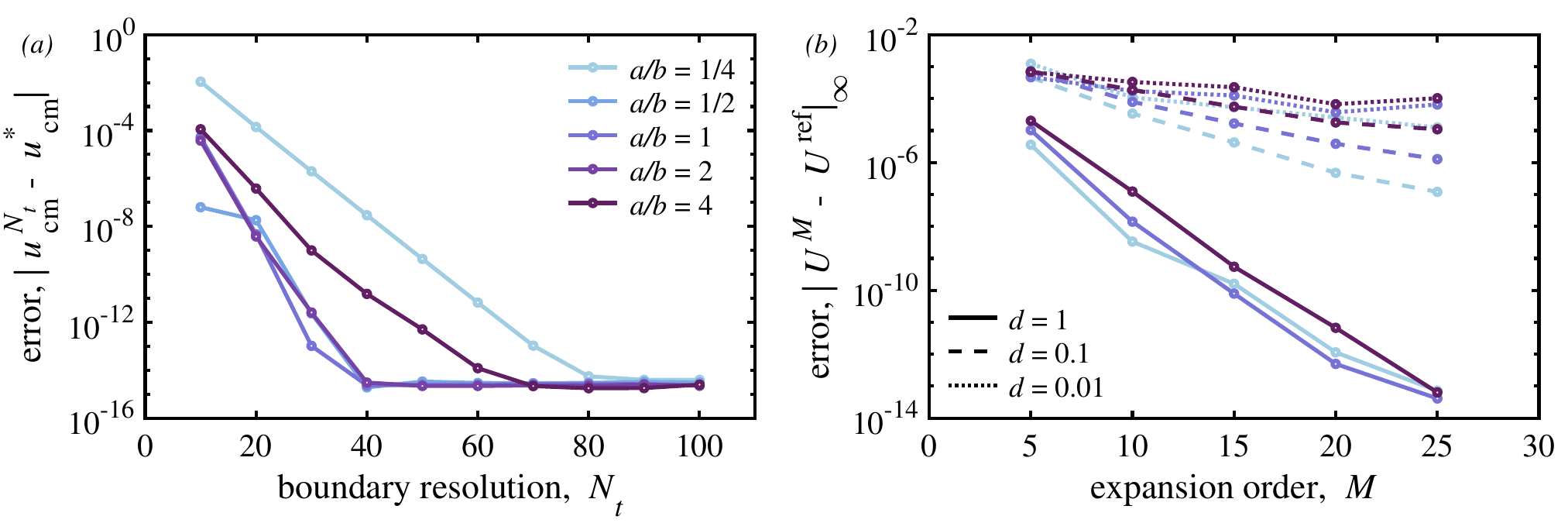}
    \caption{Convergence of {\em(a)} the single-particle solver against the analytical center-of-mass velocity for various particle aspect ratios and {\em(b)} a pair of close particles. The former exhibits exponential convergence in the number of arclength nodes $N_t$. We also observe exponential convergence in $M$ for the pair simulations, but the rate of convergence decreases significantly with the edge-to-edge separation distance.}\label{fig:convergence}
\end{figure}

We assess the convergence of our numerical method against two problems: the analytical expression for the translational velocity of a single particle \eqref{eq:ucm-exact}, and the translational velocity of a pair of close particles. The former is determined entirely by the first mode of $\sigma$, meaning only the $M = 1$ interior mode is needed, and thus provides a test to isolate the surfactant solve. Figure \ref{fig:convergence}{\em(a)} shows the results of these computations. We observe exponential convergence for all aspect ratios, reaching near machine precision with $N_t = 80$ points. High aspect ratio particles naturally require higher resolution, with $a/b=0.25$ and $a/b =4$ requiring roughly twice as many points compared to the circular case $a/b = 1$. The pair refinement study on the interior (fixed $N_t=1024$) is shown in panel {\em(b)} for aspect ratios $a/b=0.25,1,4$ and edge-to-edge separation distances $1, 0.1, 0.01$. We also observe approximately exponential convergence in $M$, however the rate of convergence decreases significantly with the separation distance. In the most challenging case, $a/b=4$, $d=0.01$, we achieve about four digits of accuracy. This accuracy is limited by the expansion order $M$, which is unable to resolve the sharp force density near the edges of closely spaced particles.

\end{appen}

\bibliographystyle{jfm}
\bibliography{refs}

\end{document}